\documentstyle[12pt,aasms4]{article}

\def\wig#1{\mathrel{\hbox{\hbox to 0pt{%
          \lower.5ex\hbox{$\sim$}\hss}\raise.4ex\hbox{$#1$}}}}

\def\etal{{\it et~al.\,}}
\def\mo{$M_\odot$}
\def\lo{$L_\odot\,$}
\def\mj{$M_{\rm J}\,$}

\def\Dwa{$\,$\uppercase\expandafter{\romannumeral5}$\,$}
\def\mic{$\mu$m$\,$}
\def\undertext#1{$\underline{\smash{\hbox{#1}}}$}
\def\sles{\lower2pt\hbox{$\buildrel {\scriptstyle <}
   \over {\scriptstyle\sim}$}}
\def\sgreat{\lower2pt\hbox{$\buildrel {\scriptstyle >}
   \over {\scriptstyle\sim}$}}
\def\sharpnull#1{}
\def\aa{Astron. Astrophys.\ }

\lefthead{Burrows \etal}
\righthead{Chemical Equilibrium Models}
\slugcomment{Submitted to the Ap. J.}
\slugcomment{University of Arizona Theoretical Astrophysics Program preprint ??-??}
\slugcomment{\today}

\begin{document}

\title{Chemical Equilibrium Abundances in Brown Dwarf and Extrasolar Giant Planet Atmospheres}

\author{Adam Burrows\altaffilmark{1} \& C.M. Sharp\altaffilmark{1}}

\altaffiltext{1}{Department of Astronomy and Steward Observatory, 
                 The University of Arizona, Tucson, AZ \ 85721}

\begin{abstract}

We calculate detailed chemical abundance profiles for a variety of brown dwarf
and extrasolar giant planet atmosphere models, focusing in particular on
Gliese 229B, and derive the systematics of the changes
in the dominant reservoirs of the major elements with altitude and temperature.
We assume an Anders and Grevesse (1989) solar composition of 27 chemical
elements and track 330 gas--phase species, including the monatomic 
forms of the elements, as well as about 120 condensates.
We address the issue of the formation and composition of clouds
in the cool atmospheres of substellar objects and explore 
the rain out and depletion of refractories.  We conclude that 
the opacity of clouds of low--temperature ($\le$900 K), small--radius condensibles 
(specific chlorides and sulfides, not silicates), may be responsible for the steep spectrum
of Gliese 229B observed in the near infrared below 1 \mic.
Furthermore, we assemble a temperature sequence of chemical transitions
in substellar atmospheres that may be used to anchor and define a sequence of spectral 
types for substellar objects with T$_{\rm eff}$s from $\sim$2200 K to $\sim$100 K.

\end{abstract}


\section{Introduction}

After a decade of ambiguous detections, bona fide brown dwarfs and extrasolar 
giant plants (EGPs; Burrows \etal 1995) are now being discovered at an accelerating pace.
Whether by radial velocity techniques (Mayor \& Queloz 1995; Marcy \& Butler 1996;
Butler \& Marcy 1996; Latham \etal 1989; Cochran \etal 1997), by direct detection in the field 
(Tinney, Delfosse, \& Forveille 1997; Delfosse \etal 1998; Tinney \etal 1998 (DENIS); Kirkpatrick, 
Beichman, \& Skrutskie 1997; Kirkpatrick \etal 1998 (2MASS)),
by direct detection in young stellar clusters (Zapatero-Osorio, Rebolo, \& Martin 1997; Comeron \etal 1993), 
by direct detection around nearby stars (Nakajima \etal 1995; Oppenheimer \etal 1995; Matthews \etal 1996;
Oppenheimer \etal 1998 (Gl 229B)), or from space (Terebey \etal 1998 (NICMOS)), the expanding census of objects
beyond the solar system with masses between 0.5 \mj and 80 \mj is lending new urgency to 
theoretical efforts to understand their evolution, spectra, and 
compositions.\footnote{\mj~ is the mass of Jupiter and is equal to $\sim 9.55\times10^{-4}$ \mo.} 
At the low temperatures (100 K $\le$ T $\le$ 2500 K) achieved in the dense, high--gravity atmospheres
of brown dwarfs and EGPs, chemical species not encountered in traditional stellar realms
assume a new and central importance.  

The molecular compositions of these exotic, low--ionization atmospheres 
can serve as diagnostics of temperature, mass, and elemental abundance and can help define
a spectral sequence, just as the presence or absence of spectral features associated with various
ionization states of dominant, or spectroscopically active, 
atoms and simple molecules does for M through O stars.  
However, the multiplicity of molecules that appear in their atmospheres lends an additional complexity to 
the study of substellar mass objects that is both helpfully diagnostic and confusing.  
Nowhere is the latter more apparent than in the appearance 
at low temperatures of refractory grains and clouds.
These condensed species can contribute significant opacity 
and can alter an atmosphere's temperature/pressure
profile and its albedo.  Grain and cloud droplet opacities 
depend upon the particle size and shape distribution
and these are intertwined with the meteorology (convection) in complex ways.  
Furthermore, condensed species can rain out and deplete the upper atmosphere of heavy elements, thereby 
changing the composition and the observed spectrum.
Hence, in brown dwarf and EGP atmospheres, abundance and temperature/pressure profiles, 
particle properties, spectra, and meteorology are inextricably linked.

One might naturally throw up one's hands at the messiness of this problem were it not
for two things: 1) there are chemical systematics that somewhat simplify the study
of these atmospheres and 2) further progress in understanding the edge of the main sequence,
brown dwarfs, and EGPs hinges directly upon its resolution.  
The goal of this paper is to explore and clarify the
chemical condensation sequences, molecular abundance profiles, and molecular spectral
diagnostics of brown dwarf and EGP atmospheres.   

The formation of refractory silicate grains
below 2500 K was already shown by Lunine \etal (1989) and Burrows \etal (1989)
to influence the evolution of late M dwarfs and young brown dwarfs through their ``Mie'' opacity.
The blanketing effect they provide lowers the effective temperature (T$_{\rm eff}$)
and luminosity (L) of the main sequence edge mass from about 2000 K and $10^{-4}$ \lo
to about 1750 K and $6\times10^{-5}$ \lo, an effect recently verified by Chabrier \etal (1998).
In addition, grain opacity slightly delays the cooling of older brown dwarfs, imprinting a slight
bump on their luminosity/age trajectories (see Figure 7 of Burrows \etal 1997).
The presence of grains in late M dwarf spectra was invoked 
to explain the weakening of the TiO bands and the shallowing of their H$_2$O troughs in the
near infrared (Tsuji \etal 1996; Jones \& Tsuji 1997).  Tsuji 
and collaborators concluded that titanium was being depleted into refractories,
a conclusion with which we agree (see \S 3).

Gl 229B (Nakajima \etal 1995; Oppenheimer \etal 1998) is a Rosetta stone for brown dwarf research.
With a T$_{\rm eff}$ of $\sim$950 K, a luminosity below  $10^{-5}$ \lo, and spectra or photometry
from the $R$ Band through 5 \mic, Gl 229B hints at or exemplifies all of the unique characteristics
of the family (Marley \etal 1996; Allard \etal 1996): metal (Fe, Ti, V, Ca, Mg, Al, Si) depletions, 
the dominance of H$_2$O vapor, the appearance of CH$_4$ and alkali metals, and the signatures of clouds.
Clouds of low--temperature condensible species above the photosphere are the most
natural explanation for the steep drop below 1 \mic in the Keck spectra between 0.83 \mic
and 1 \mic (Oppenheimer \etal 1998).  These clouds may not be made up of the classic
silicate refractories formed at much higher temperatures, since these species 
have probably rained out (Marley \etal 1996).  From simple Mie theory, their mean particle size must be 
small ($\sim$ 0.2 \mic) in order to influence the ``optical'' without much perturbing the near infrared.
In addition, such a population of small droplets can help explain why 
Gl 229B's near--infrared troughs at 1.8 \mic and 3.0 \mic are not as deep as theory would otherwise
have predicted.  Just as Tsuji and collaborators have shown that silicate grains at higher temperatures 
can shallow out the H$_2$O troughs, so too can species that condense at lower temperatures
($\le$ 1000 K ?) explain the shallower--than--predicted Gl 229B H$_2$O troughs.  What those species might be
can be illuminated by chemical abundance studies and is one of the subjects of this paper.
Note that a cloud grammage in these small--radius low--temperature refractories of only $\sim 10^{-5}$ gm cm$^{-2}$ 
would be adequate to explain the anomalies. 

With Gl 229B as a reference benchmark, we explore the composition profiles
of low--temperature brown dwarfs and EGPs to derive the systematics of the changes
in the dominant reservoirs of the major elements (Table 1).  These equilibrium chemical sequences
are predominantly a function of temperature and can help to define a spectral sequence
for substellar objects from the main sequence edge near 2000 K to EGPs with T$_{\rm eff}$s of a few
hundred Kelvin.  The appearance and disappearance of various molecules and refractories delineates
an effective temperature sequence and the new proposed ``L'' dwarf 
spectral classification (Kirkpatrick \etal 1998) may correspond to a subset of the compositional 
sequences we derive.

However, as mentioned above, the full problem requires that the composition profiles, opacities,
meteorology, temperature/pressure profiles, and spectra be handled self--consistently.
At the present time, given the ambiguities in the grain/cloud properties and the paucity
of optical constants, such a program is not realistic.  In lieu of this, we employ the
temperature/pressure profiles derived in Burrows \etal (1997).  This allows us to define
the overall trends, while at the same time focusing on the essential chemistry.
The models of Burrows \etal (1997) consistently incorporate the abundances of the 
major gas--phase species, with their opacities, but do not include the thermal effects of the
low--temperature condensibles.  Nevertheless, the thermal profiles below 
the photospheres (at higher pressures) are always close to adiabatic,
while the thermal profiles at lower optical depths are generally super--adiabatic ($|dT/dr|$ small).
Coolants in the atmosphere and non--gray effects generally lower the temperatures achieved
at the lower pressures far below the Milne temperature ($\sim$ 0.81T$_{\rm eff}$) to 
near $\sim$ 0.5T$_{\rm eff}$.  This has the consequence that species that would
not otherwise condense in Gl 229B--like atmospheres may in fact be present.
The Burrows \etal (1997) models incorporate such cooling and non--Eddington
effects.   In \S 2, we describe the techniques and procedures we employ to
perform the equilibrium calculations.  In \S 3, we present  
temperature profiles for representative model brown dwarf and EGP atmospheres.   
We also calculate various composition boundaries in temperature--pressure space 
between the major compounds of the abundant elements.  In addition, 
element by element we describe the abundance profiles in realistic brown
dwarf/EGP model atmospheres.  This is the central section of the paper.  
In \S 4, we explore the potential effects of
rainout and depletions on the remaining mix of dominant chemical species.  
In \S 5, we summarize the list and order, as a function of temperature, of
chemical indicators that can help to define a spectral classification scheme
for substellar objects and we review our general findings.

\section{Techniques, Algorithms, and Procedures}

Using the brown dwarf and giant planet model atmospheres of Burrows {\it et al.}~(1997),
and assuming an Anders and Grevesse (1989) solar composition of the 27 chemical
elements H, He, Li, C, N, O, F, Ne, Na, Mg, Al, Si, P, S, Cl, Ar, K, Ca, Ti, V,
Cr, Mn, Fe, Co, Ni, Rb and Cs, we calculate the abundances of each of about 330
gas--phase species, including the monatomic forms of the elements, and of about 120
specified condensates.  These calculations are performed at each of about 2000
temperature-pressure points for every atmosphere model.

Table 1 lists the abundances by number for the
27 elements used in these calculations.  With the exception of lithium, rubidium
and cesium, the list covers nearly all the most abundant elements anticipated to be
in brown dwarf atmospheres.  Lithium is added because it is an important brown dwarf
indicator.  Rubidium and cesium are added because they have recently been detected
in cool objects (Oppenheimer \etal 1998; Tinney \etal 1998).  
Their very low excitation and ionization potentials make
them good temperature indicators in a range were many other elements are associated
into molecules or have condensed out.

The same methods were used as in Sharp and Huebner (1990), but the improved
version of the original computer code SOLGASMIX (Besmann 1977) used in that
paper was further improved in the current work by including more species,
incorporating a number of error traps, and adding
features to make the code more user-friendly.  Using the multi-dimensional
Newton-Raphson method of White, Johnson and Dantzig (1958), the equilibrium
abundances of gas--phase and condensed--phase species at a given temperature,
pressure, and elemental composition were obtained by minimizing the total free
energy of the system.  After subtracting the constraint equation
in  the manner of Lagrange's method of undetermined multipliers and
linearizing, SOLGASMIX uses Gaussian elimination with pivoting to iteratively
solve the resulting matrix equations.  The free energy of each species at a
particular temperature is obtained by evaluating a polynomial of the form

\begin{equation}
\Delta G_{\phi i}(T) = \frac{a}{T} + b + cT + dT^2 + eT^3,
\label{eq:fit}
\end{equation}
where $a$, $b$, $c$, $d$ and $e$ are fitted coefficients and
$\Delta G_{\phi i}(T)$ is the Gibbs free energy of formation at temperature $T$
of species $i$ in phase $\phi$, relative to its consitituent atoms in their
monatomic neutral gaseous state.  For instance, for H$_2$S the reference states
of the elements are monatomic hydrogen and sulfur in the gas phase, whose
energies of formation are zero by definition.  If the species being considered
is in the gas phase and is monatomic, all the coefficients on the right hand
side of eq. (\ref{eq:fit}) are zero.

In the original work of Sharp and Huebner (1990), the total free energy of the
system was expressed quite generally in terms of the gas phase, several
liquid or solid solutions (each containing at least two species), and distinct
condensed phases of invariant stoichiometry (each containing a single species).
In our current work, because of the large number of species considered, we
decided to simplify the calculations and omitted solutions, such as the
solid solution of melilite (Ca$_2$Al$_2$SiO$_7$ with Ca$_2$MgSi$_2$O$_7$).
Thus, given a value of $\Delta G_{\phi i}(T)$ from eq. (\ref{eq:fit}) for each
chemical species, the elemental composition, and initial trial
values of the species abundances, the total equilibrium free energy of the
system is found by minimizing

\begin{equation}
\frac{G(T)}{RT} = \sum_{i=1}^{m} \left[n_{\phi i} 
\left\{ \frac{\Delta G_{\phi i}(T)}{RT} + \ln P + \ln (\frac{n_{\phi i}}{N}) \right\}
\right]_{\phi=1} + \frac{1}{RT} \sum_{\phi=2}^{s+1} 
\Big[n_{\phi i} \Delta G_{\phi i}(T)\Big]_{i=1},
\label{eq:minig}
\end{equation}
where $R$ is the gas constant, $P$ is the total pressure in atmospheres, and
$N$ is the total number of moles in the gas phase.  The first sum is
over $m$ species in the gas phase with $\phi=1$, the second sum is over $s$
condensed phases (numbered 2 to $s+1$), and $n_{\phi i}$ is the number of
moles of species $i$ in phase $\phi$.  Each condensed phase is in fact only
a single species, so $i=1$ in the second sum, since we need to distinguish
between, for example, H$_2$O(gas) and H$_2$O(liquid), which are considered
separate species.  The subsidiary mass balance relations are satistfied for
each element $j$ of the total number of elements, $k$, as follows:	

\begin{equation}
\sum_{i=1}^{m}      \Big[ \nu_{\phi ij} n_{\phi i} \Big]_{\phi=1} +
\sum_{\phi=2}^{s+1} \Big[ \nu_{\phi ij} n_{\phi i} \Big]_{i=1}    = b_j
\mbox{ ~~~for {\it j = 1} to {\it k} ,}
\label{eq:conserve}
\end{equation}
where $\nu_{\phi ij}$ is the stoichiometric coefficient of element $j$ in
species $i$ in phase $\phi$ and $b_j$ is the number of gram-atoms of element
$j$.  As in eq. (\ref{eq:minig}), the first sum is over the gas-phase species
($\phi=1$) and the second sum is over all the single component ($i=1$)
condensates.

Most of the thermodynamic data were obtained from the JANAF tables (Chase
1982; Chase {\it et al.} 1985).  Data on a number of condensates not available
in these tables were obtained from Turkdogan (1980) and data on the two
condensates NaAlSi$_3$O$_8$ (high albite) and KAlSi$_3$O$_8$ (high sanidine)
were obtained from Robie and Waldbaum (1968).  Tsuji (1973) was the source
of data for the gas-phase molecules CaH, CrH, MnH, NiH, MnO, NiO, MnS, TiS, TiN,
SiH$_2$, SiH$_3$, together with some carbides of little importance in the brown
dwarf context.  Our database currently holds information on 1662 species, of which
about 450 are followed in these calculations.  For those species for which the
JANAF tables were used, the coefficients in eq. (\ref{eq:fit}) were obtained from
a database generated in Sharp and Huebner (1993), in which the free energies were
renormalized relative to the monatomic gaseous phase of each element, rather
than relative to standard reference states.  A polynomial with up to five
coefficients as in eq. (\ref{eq:fit}) was fitted over the tabulated temperature
range of each species.  Using the redefined free energies has the advantage
of removing any discontinuities associated with phase changes in the reference
states, permitting polynomial fits to be made over large temperature ranges.

In Sharp and Huebner (1990), the only alkali elements considered were sodium
and potassium, together with their most important compounds.  In this work, we
include all the other alkali elements, except francium, together with their
most important compounds, in particular their chlorides and fluorides, in both
gas and condensed phases.  With the exception of rubidium, all the
data were obtained from the JANAF tables.  As no useful data on rubidium and
its compounds are available in the JANAF tables, including in the more recent
compilations in Chase {\it et al.} (1985), we used Barin (1995).

The JANAF tables are deficient in a number of important
volatile condensates, the most important being
H$_2$O(ice).  A prescription for water from Eisenberg and Kauzmann (1969)
was modified and incorporated into our code.  A second important volatile
condensate is NH$_3$(ice).  The pure liquid phase of NH$_3$, together with
a number of its hydrated liquid phases, were not considered.  The data for
NH$_3$ were obtained from the CRC Handbook (Weast and Astle 1980) in the
form of vapor pressure coefficients.  Another important volatile
condensate is ammonium hydrogen sulfide, NH$_4$SH, which is believed to be
an important component in the Jovian atmosphere (Lewis 1969).  As with
NH$_3$(ice), we obtained the data for NH$_4$SH from the CRC Handbook,
in the form of vapor pressure coefficients.  However, unlike either
NH$_3$ or H$_2$O, which on vaporization remain the same compound,
NH$_4$SH decomposes into the two gaseous components, NH$_3$ and H$_2$S.
In addition, we added NH$_4$Cl, which decomposes into NH$_3$ and HCl, but
this species proves to be unimportant at the temperatures and pressures
encountered in giant planet and brown dwarf atmospheres.  Except for the
data originating from Tsuji (1973) and our newly added data on the
condensed phases of H$_2$O, NH$_3$, NH$_4$SH, and NH$_4$Cl, which were
expressed in different forms, all data were converted to the same form as
eq. (\ref{eq:fit}), before the evaluation of eqs. (\ref{eq:minig}) and
(\ref{eq:conserve}).  Furthermore, we replaced the polynomial fit for the
gaseous species CO found in the JANAF tables with a polynomial fit based
on its spectroscopic constants, as the spectroscopic data are very well
known for this molecule (Sharp 1985).

We assume that the gas is ideal, {\it i.e.} that the activity of each
species is the same as the partial pressure in atmospheres and, with
the exception of liquid H$_2$O, only solid condensates are included in the
calculations.  We consider the neglect of the liquid phases to be quite
reasonable, as the difference between the free energies of formation of
the liquid and solid phase of a species is generally much smaller
than the difference between the free energies of one of the condensed
phases and the gas phase.  For example, for iron in the range 1000 to
2000 K, the change in the free energy when converting the vapor into the
liquid is at least five orders of magnitude larger than the
corresponding change when converting the liquid to the solid.  In the
case of TiO, the ratio of the change in the free energy between vapor and
liquid to between liquid and solid is much smaller, but is still at least
one order of magnitude in the same range (though in detailed
calculations other condensed oxides form).  Additionally, the inclusion
of ionization slows down the calculations substantially in any temperature
region which involves both marginal ionization and condensation, as is the
case in this work, and since ionized species play a negligible role 
at the lower temperatures of substellar atmospheres, we omit them.

The main object of this work is to derive the detailed equilibrium
abundance profiles for models of the atmospheres of cool brown dwarfs and
giant planets, given their temperature-pressure profiles, with particular
reference to the brown dwarf Gliese 229B (Nakajima {\it et al.} 1995).
Since we start with temperature-pressure profiles with specified
gravities and effective temperatures from Burrows {\it et al.} (1997),
to obtain such profiles for given ages and masses, we interpolate
using the evolutionary calculations from that work.
Given a profile corresponding to the require model, we
calculate the abundances by minimizing the free energy of the system.
Starting at the high temperature end of the profile, the code uses
approximate trial values based on the element abundances to start the
calculations.  Upon convergence, the abundances of selected species
are written to a file for later graphical treatment.  These abundances
are then used as trial values for the next calculation 1~Kelvin lower
in temperature, and at selected temperature intervals, typically 100 K,
additional data are printed out.  This process is repeated for
progressively lower temperatures and pressures in the atmosphere, until
the end of the data file is reached.  At the high temperature end, 
convergence is usually rapid, since the gas consists of atomic and simple
molecular species, and few, if any, condensates.  At progressively lower
temperatures, as more species condense out and the chemistry becomes
more complicated, convergence is slower and more ``delicate.''  If the
starting values before each iteration are too far away from the solution,
there is a risk of non-convergence.  It is for this reason that the
calculations are always started at the high temperature end of a profile.

However, even with the above method, convergence problems can arise.  It is
found that between 1100 and 1200 K for pressures less than about 0.1
atmospheres the program may fail to converge.  For higher pressures, the code
generally operates successfully until temperatures below 600 K are reached.
In the first case, non-convergence is often associated with the refractory
condensate Ca$_2$MgSi$_2$O$_7$, and in the second case it is often
associated with the less refractory (and very low abundance) condensate
LiAlO$_2$.  In both cases, these condensates appear at higher temperatures,
only to disappear into other species before the problems are encountered.
Upon removal of these and many other similar species, such as Al$_2$O$_3$,
which is present only above about 1800~K in most models and disappears into
other aluminum-bearing species at lower temperatures, the problems are often
solved.  These problems are caused by a too rapid change in
abundance associated with condensation or a phase change.  Unfortunately, it
is often not possible without doing the calculations to know {\it ab initio}
which species are likely to be unimportant and can be neglected.
Fortunately, removing gas-phase species of negligible abundance can speed up
the calculations.  After removing redundant species, iterations can be restarted.

If it was still not possible to obtain successful convergence after removing
one or more redundant species, condensates that are present, but that can be
considered inert, can be removed from the calculations, which can be
restarted with trial values from a previous dump file.  In the
process of removing a condensate from the calculations, the corresponding
abundances of each element tied up in the species is removed.  For example,
if the condensate spinel, MgAl$_2$O$_4$, is present and is to be removed
from the calculations, then given N(Mg) and n(MgAl$_2$O$_4$), which are,
respectively, the number of gram-atoms of Mg in all forms in the calculations
and the number of moles of spinel, the corrected total abundance of Mg is
N(Mg)--n(MgAl$_2$O$_4$), that of Al is N(Al)--2n(MgAl$_2$O$_4$), and that
of O is N(O)--4n(MgAl$_2$O$_4$), from which the relative element abundances
can be obtained. 

However, it was often found that at a lower temperature, removing a
condensate changed the chemistry.  A very good example is iron, which in
our calculations normally forms the solid metal at high temperatures.
If iron is retained in the calculations, at 718~K iron metal reacts with
H$_2$S to form FeS, which is then the main compound of sulfur, and
substantially depletes sulfur from the gas phase.  If iron is removed
from the calculations when it initially condenses, much more sulfur
remains in the gas phase, which is then available to form the volatile
condensate, NH$_4$SH, at lower temperatures.  Hence, the presence of
NH$_4$SH clouds inferred in Jupiter's atmosphere may be an indirect
consequence of the rainout of iron metal to deeper regions.

\section{Composition Boundaries and Abundance Profiles}

Figure 1 depicts temperature--pressure profiles for 5, 10, 20, and
40 \mj atmosphere models from Burrows \etal (1997), for ages of 0.1, 1.0, and 10.0 Gyrs.
Only the 5 and 10 \mj models are given at 0.1 Gyr.  The dots identify
the positions of the photospheres.  This set of profiles represents the
general trends in brown dwarfs and EGPs.  Note that the concept of a photosphere
is ambiguous, since the opacities are stiff functions of wavelength.  Radiation
at different wavelengths decouples from a variety of disparate pressure levels.
By photosphere we mean here the level at which the temperature equals the effective
temperature, defined via Stefan's Law.  This approach provides us with a sensible 
average position, but should not be overinterpreted.  

Higher--mass models generally reside
at higher entropies, {\it i.e.}, at a given temperature level the gas is at lower 
pressures.  However, as Figure 1 demonstrates, cooling in the 
radiative zones above the photosphere at lower temperatures can 
result in a reversal of this behavior.  Furthermore, for a given mass the older a substellar
object the lower the effective temperature and the higher the photospheric pressure.
In addition, for a given age the higher the mass the higher the photospheric pressure.
However, and importantly, although the effective temperature is a strong function
of age and mass, the corresponding pressure at the photosphere is only weakly
dependent upon these quantities, and varies within $\sim$ half a dex of 1 atmosphere.
Consequently, the abundances of species at the photosphere 
of a brown dwarf or EGP are governed almost entirely by
the temperature.  This is an important conceptual simplification.

Figures 2 through 4 depict the regions of transition from one chemical
species to another in temperature--pressure space for a variety of important
constituents of equilibrium substellar atmospheres.   Anders \& Grevesse (1989)
solar abundances are assumed (Table 1).  Only a subset
of the calculated and interesting composition boundaries are depicted.
Nevertheless, these figures summarize some of the diagnostic chemical sequences
and transitions encountered in theoretical substellar atmospheres and the 
reservoirs of the major elements.  Superimposed on these plots are 
atmospheric profiles of 5, 10, 20, and 40 \mj models at 1 Gyr from Figure 1.  
Also shown for comparison on Figures 2 through 4 are adiabats above and below the photospheres
for more massive M dwarf models of Burrows \etal (1993) at 0.08, 0.09, and 0.115 \mo$\,$ and 1 Gyr.  
As demonstrated in Figure 1, for a broad range of objects 
and ages, the narrow band around 1 atmosphere is relevant
when looking at the compositions near the photosphere.  
Furthermore, the near verticality of the transition
curves confirms that composition is most closely linked with temperature.

Figures 5 through 14 depict theoretical chemical equilibrium composition profiles
versus temperature for Gl 229B--like models (see Marley \etal 1996; Allard \etal 1996) as a function of  
T$_{\rm eff}$ and gravity. To avoid confusion, only a subset of the major compounds
are included in the plots.  Implicit are the corresponding pressures 
from the atmosphere models of Burrows \etal (1997).
The odd--numbered figures depict the gravity dependence for T$_{\rm eff}$ = 950 K
and the even--numbered figures depict the T$_{\rm eff}$ dependence for 
a gravity of 1000 m s$^{-2}$.  The figures are approximately in order of
decreasing elemental abundance. Hence, Figures 5 through 14 and the curves 
on those figures follow the rough sequence: O, C, N, Mg, Si, Fe, S, Al, Ca, Na, P, K, 
Ti, V, Li, Rb, and Cs.  We will describe each of these figures in turn,
but even though they depict the range of theoretical abundance profiles
for Gl 229B models, they demonstrate that the sequence of chemical reservoirs of the major elements
as a function of temperature varies little with model.   This is a consequence of the
weak pressure and gravity dependence of the equilibrium 
abundances and the steepness of the 
transition boundary curves in Figures 2 through 4 
relative to the model temperature--pressure profiles.
Hence, the sequence of dominant chemical species for an
element along the entire brown dwarf/EGP continuum and the transition 
temperatures from one species to another are very approximately universal.
Note that the proximity of a species to the photosphere and
its importance in the emergent spectrum does vary importantly
as a function of age and mass or gravity and T$_{\rm eff}$.
Tables 2a and 2b provide useful lists of the major chemical species of
each of the most abundant elements found in the atmospheres of
substellar objects.

\subsection{Oxygen, Carbon, Nitrogen, Iron, Chromium, Sulfur, Titanium, Vanadium}

As Figures 5 and 6 demonstrate, the major reservoir for 
oxygen in substellar atmospheres is gaseous H$_2$O.
The only significant competitor for abundant oxygen is
CO, but CO is converted into CH$_4$ between temperatures of 1100 and 2000 K.
Only at temperatures below $\sim$300 K does H$_2$O condense,
as the liquid above 273.15 K, otherwise as ice. 
As is well known (Fegley \& Lodders 1996), Jupiter 
has condensed H$_2$O below its photosphere.
As is not so well known and as is demonstrated in Figure 2, 
at 1 Gyr a 10 \mj object may have condensed H$_2$O in its atmosphere.
In fact, as Figure 2 suggests, all but the youngest EGPs and those
not at elevated temperatures due to proximity to a primary ({\it e.g.},
51 Peg b, $\tau$ Boo b) should have water clouds (Burrows \etal 1997).
This may importantly affect their albedos and reflected spectra
(Marley \etal 1998).

As alluded to above and as shown in Figures 5 and 6, 
the major reservoirs of carbon are CO and CH$_4$.  The reaction 
$$ CO + 3H_2 \rightarrow H_2O + CH_4$$ and Le Chatelier's Principle
indicate that for a given temperature CO is favored at low pressures
and gravities.  CO is also favored at high temperatures and for
models with high T$_{\rm eff}$s.  From Figures 5 and 6, it is expected
that in Gl 229B CO will convert into CH$_4$ between temperatures of 
1300 and 1800 K, depending upon T$_{\rm eff}$ and gravity.  To put
equilibrium CO closer to the photosphere in Gl 229B, high T$_{\rm eff}$s and 
lower gravities are preferred.  This makes the interpretation
of the detection in Gl 229B of CO at 4.67 \mic (Noll, Geballe, \& Marley 1997) 
problematic, since CO is not expected in great
abundance near the photosphere for T$_{\rm eff}$s near 950 K.
This may imply that the higher CO abundances at the higher temperatures at depth
are being advected out of equilibrium into the cooler regions of the photosphere
and that the kinetics of the reaction might have to be considered. 
Be that as it may, CH$_4$ should be the major atmospheric reservoir of carbon for all but
the youngest and hottest (T$_{\rm eff}$ $\ge$ 1300 K) brown dwarfs and EGPs.

Nitrogen is in the form of N$_2$ for temperatures above 800 K,
but is converted into NH$_3$ between 600 and 800 K (Figures 5 and 6).
From Figure 2, we see that the NH$_3$/N$_2$ boundary coincides with 
the typical atmospheric pressure level ($\sim$ 1 atmosphere) near
the $Fe \rightarrow FeS$ transition. This flags the $\sim$700 K level
in a substellar object as another interesting transition.
From Figures 2, 7 and 8, we see that for iron the first condensate to 
form (at $\sim$ 2200 K) is the metal, followed at much
lower temperatures by the solid sulfide, FeS.  FeH should also be present in abundance, but is not
in our calculations (Fegley \& Lodders 1996).  The  
major reservoir for sulfur above 718 K is gaseous H$_2$S. As Guillot \etal (1997)
have shown, at lower temperatures ZnS forms.  Note that though Na$_2$S (c) and K$_2$S (c)
are in our calculations, they never appear in great abundance (see \S 3.2).  FeS
always forms at 718 K due to the reaction of iron metal with H$_2$S.
The pressure--independence of this reaction follows
from simple mass action arguments applied to the reaction
$$Fe (c) + H_2S \rightarrow H_2 + FeS (c)$$
and assumes that metallic iron is
available in the equilibria to react with the buffer gas H$_2$S.

At $\sim$2000 K, chromium exists predominantly in the form of gaseous CrH, 
but is converted in large part into
the metal near 1560 K ($\pm 15$ K, depending upon the pressure (gravity)).
Near and below 1500 K, the dominant compound of chromium is solid Cr$_2$O$_3$.
At 1500 K, CrH exists, but involves only a few percent by number of the available 
element.  Hence, the disappearance of CrH (g) occurs around 1400--1500 K. 

At temperatures below $\sim$200 K, NH$_3$ condenses into clouds.
Such clouds are seen in Jupiter and Saturn, but as implied in Figure
2 they may be present in the lowest--mass EGPs with T$_{\rm eff}$s below
$\sim$300 K.

The ratio \{Ti$_{\mbox{con}}$\}/TiO on Figure 2 indicates where the abundance of one or
more titanium condensates is equal to that of the gas--phase TiO, since the first
titanium condensate to form with decreasing temperature depends upon the
pressure.  Note that for pressures corresponding to red dwarf atmospheres,
perovskite, CaTiO$_3$, is the first titanium condensate to appear, whereas
for sub-stellar objects this is replaced by Ti$_3$O$_5$.  
As indicated in Figure 2, there is a range
of pressures where both form.  At lower temperatures, the first
titanium condensate to form changes into other compounds (Ti$_3$O$_5$, Ti$_2$O$_3$
Ti$_4$O$_7$, and MgTi$_2$O$_5$). Equilibrium abundance profiles for these species
for various models of Gl 229B are shown in Figures 11 and 12.
The Ti transition sequence starting from $\sim$2400 K to $\sim$2100 K
is quite general for substellar objects with T$_{\rm eff}$s below $\sim$2000 K.
For T$_{\rm eff}$s above $\sim$2000 K, perovskite replaces 
Ti$_3$O$_5$ as the first condensed species
into which TiO is transformed and titanium is depleted.

The main gas--phase species of vanadium and the first condensate to
appear are both VO, and the curve on Figure 2 labeled by \{VO\}/VO shows where these
two phases have equal abundance. The condensed phase of VO first appears
to the left of the curve.  Abundance profiles for V compounds
for a range of Gl 229B models are depicted in Figures 11 and 12.
VO is depleted into solid VO, then into V$_2$O$_3$, but its depletion sequence
starts near 1800 K at lower temperatures than for titanium.  Hence, we expect that in brown dwarfs
and EGPs (and presumably in M dwarfs) first the abundance of gaseous TiO 
will decrease (near T$_{\rm eff}$s of 2100 to 2400 K) and then (near 1800 K) 
the abundance of gaseous VO will decrease due to its depletion into solid VO. 
This depletion order should manifest itself in the spectral sequence of cool objects.

\subsection{Silicates, Phosphorus, Sodium, Potasium}

In Figure 3, the condensates MgSiO$_3$ (enstatite), Mg$_2$SiO$_4$ (fosterite),
and MgAl$_2$O$_4$ (spinel) are shown and are the most important condensates of
magnesium, silicon, and aluminum.  In the equilibrium calculations, these
condensates are found to persist to lower temperatures, and unlike 
those of titanium and vanadium, do not change into other compounds.  Figures 7 and 8
depict the equilibrium abundance profiles of enstatite and fosterite for the range
of Gl 229B models.  Figures 9 and 10 show the same for the more refractory silicates.  Also included
in Figures 3, 9, and 10 are the two main high--temperature condensates of sodium and potassium, 
NaAlSi$_3$O$_8$ (high albite) and KAlSi$_3$O$_8$ (high sanidine).

As indicated on Figures 9 and 10, the most refractory compounds are the calcium
aluminum silicates ({\it e.g.}, akermanite), followed by the calcium magnesium (iron) silicates.  These condensates can be
found from temperatures of $\sim$2300 K down to $\sim$1800 K.  Above $\sim$2000 K, a 
major reservoir of Si is gaseous SiO (not shown in the figures), which is 
progressively sacrificed to form the refractories as the temperature decreases.  Below 1800 K, 
among the classical refractories only spinel, enstatite, fosterite, 
and diopside (CaMgSi$_2$O$_6$) remain in abundance.
In equilibrium, diopside is the low--temperature reservoir of calcium, while that of 
aluminum transitions from spinel to high albite back to spinel as the temperature decreases.
The latter transition occurs below the minimum temperature shown in Figures 9 and 10.
The aluminum--bearing compound, corundum (Al$_2$O$_3$), is not a major component of 
brown dwarf atmospheres.
As Figure 3 demonstrates, enstatite and fosterite appear near the photospheres
of brown and M dwarfs at temperatures from 1700 to 1900 K.  In fact, as the relevant figures
imply, classical silicate grains form in a temperature range from 2400 K to 1500 K.  This is just the T$_{\rm eff}$ transition
regime between stars and brown dwarfs and serves to emphasize the importance of grain opacity
in determining the properties of transition objects and in setting the T$_{\rm eff}$ and luminosity
of the main sequence edge. Its role has yet to be precisely delineated, but Burrows \etal (1993)
note that at the edge T$_{\rm eff}$ is decreased by $\sim$250 K to $\sim$1750 K when the approximate effects
of grains are included.

At high temperatures, sodium and potassium are in monatomic form, but at temperatures of
1300 to 1500 K (Figures 3, 9, and 10) NaAlSi$_3$O$_8$ and KAlSi$_3$O$_8$ condense.
Figure 4 depicts the Na/NaCl(g) and K/KCl(g) transition boundaries that occur in the 1000 K
to 1500 K temperature range and near the photosphere for T$_{\rm eff}$s of 1200 to 1400 K.
Not shown on Figure 4, these chlorides condense below 800 K.   
Hence, as the temperature decreases, the major sodium species 
are first atomic Na, then gaseous NaCl and solid NaAlSi$_3$O$_8$, and finally solid NaCl.  Potassium shows a similar
sequence.  The presence of condensed NaCl and KCl at low temperatures (below the inferred
T$_{\rm eff}$ of Gl 229B !) is quite suggestive and we will return to this in \S 4.

Figures 9 and 10 depict representative abundance profiles for the dominant phosphorus compounds
and Figure 3 depicts the major chemical transitions phosphorus undergoes.  The PH$_3$
$\rightarrow$ P$_4$O$_6$ $\rightarrow$ Mg$_3$P$_2$O$_8$ transformations occur
in the temperature range of 800 to 1200 K.
The chemistry of phosphorus is particularly interesting, as the condensation
temperature of Mg$_3$P$_2$O$_8$ (magnesium orthophosphate) is only weakly
dependent upon pressure and forms very close to the photosphere of many of
the depicted brown dwarf models.  The predominant gas--phase
species of phosphorus at about 1200 K is PH$_3$ (phosphine).  Except at
pressures above about 30 atmospheres, PH$_3$ is rapidly oxidized by H$_2$O
to form the gas-phase oxide P$_4$O$_6$, which becomes by far the dominant
compound of phosphorus until Mg$_3$P$_2$O$_8$ condenses.  At higher
pressures, Mg$_3$P$_2$O$_8$ first condenses and then the remaining gas--phase phosphorus 
forms the oxide.  The curve on Figure 3 labeled
(P$_4$O$_6$+\{Mg$_3$P$_2$O$_8$\})/PH$_3$ indicates where the abundance of
PH$_3$ is the same as the sum of the abundances of the oxide and the
condensate, which except for the highest pressures where Mg$_3$P$_2$O$_8$
condenses first, is where the ratio P$_4$O$_6$/PH$_3$ is unity.  
Note that the phosphorus equilibrium can be represented by the reaction:
$$ 4 PH_3 + 6 H_2O \rightarrow P_4O_6 +  12 H_2\, , $$
which involves more gas--phase molecules on the 
left--hand--side than the right.   Hence, Le Chatelier's
Principle can be used to explain why the slope of the 
phosphorus curve is anomalously positive,
unlike that of the other curves in Figure 3.

However, according to Borunov, Dorofeeva, \& Khodakovsky
(1995), the thermodynamic data for
the gaseous species P$_4$O$_6$ in the JANAF tables may be in error,
and may result in an overestimate of its abundance.  In their
work, they report a less negative value for the enthalpy of formation
of P$_4$O$_6$, which would be reflected in a lower abundance.  We have
been using the data from the JANAF tables, and find that with decreasing
temperature P$_4$O$_6$ replaces phosphine, PH$_3$, as the dominant
gas-phase phosphorus-bearing compound below about 1100 K in brown dwarf
atmosphere models, until the condensate magnesium orthophosphate,
Mg$_3$P$_2$O$_8$, forms between about 800 and 900 K, as shown in Figure 3 (assuming other
magnesium condensates are available to react with phosphorus--bearing
species).  It may be that the abundance of P$_4$O$_6$ is overestimated
in our calculations.  However, for a present-day Jupiter model, and
other similar models, magnesium orthophosphate condenses near to or
above the temperature at which the phosphorus chemistry of the gas
phase switches over from being dominated by PH$_3$ to being dominated
by P$_4$O$_6$, so it appears that P$_4$O$_6$ may not in any case play
an important role.  In Figure 3, the curves indicating the
phosphorus chemistry clearly show that at high pressures the
condensation of Mg$_3$P$_2$O$_8$ overlaps with the formation of
P$_4$O$_6$.

\subsection{Lithium, Cesium, Rubidium, Chlorine, Fluorine}

Figure 4 depicts the atomic to gaseous chloride transitions for the alkali metals
and compares them to model temperature--pressure profiles from Burrows
\etal (1997) at 1 Gyr.   Figures 13 and 14 show the abundance profiles for Li,
Cs, and Rb species for various Gl 229B models.  The chemistry of these metals
is similar, but since the elemental abundances are quite different (Table 1), there
is the natural ordering (Li, Rb, Cs) depicted in the figures.  Lithium 
is in atomic form above $\sim$1700 K, but is converted first into LiOH,
then into LiCl at lower temperatures.   Below $\sim$700 K, in equilibrium
lithium resides in LiAlO$_2$. 
Hence, at sufficiently low temperatures,
the ``lithium'' test at 6708 \AA\ is of no use.  Below  T$_{\rm eff}$s of perhaps 1300 to 1600 K,
atomic lithium will be depleted and the strength of the standard lithium line
will abate.   Below temperatures of 1200 K, the abundance of atomic lithium 
will be $\sim$one hundreth of solar.  Interestingly, at these low T$_{\rm eff}$s, spectral signatures  
of the chlorides may be detectable in the mid--infrared.
The consequence of this is to nullify the standard lithium test for the majority 
of brown dwarfs and EGPs, unless they are caught early in their lives ({\it e.g.}, in young clusters)
or are ``massive'' and near the transition edge. 

Rubidium and cesium form the chloride as well, but at temperatures that are 200 to 300 K lower than does Li.
Nevertheless, the alkali metals survive in atomic
form to lower temperatures than do the other true metals.  Since they are less refractory and 
survive in monatomic form for a longer stretch of temperature, the alkali metals should manifest
themselves in the spectra of transition objects after the other metals 
({\it e.g.}, Fe, Ca, Ti, V, Mg, Al) are depleted and rainout, as 
described above and in \S 4.  Very approximately, the order of disappearance of the atom with 
decreasing temperature from $\sim$1500 K to $\sim$1200 K should be: Li, Na, Cs, Rb, K,
with Cs, Rb, and K disappearing at about the same temperature.
(Note that Na and K are depleted into high albite and sanidine as well.)
Something like this seems to be what is being observed by the 2MASS (Kirkpatrick \etal 1998) 
and DENIS (Tinney \etal 1998) collaborations in the spectra of 
newly--discovered substellar (?) and transition objects in the field.

From Figures 4, 13, and 14 we would predict that atomic cesium would be 
weak or non--existent in the spectrum of Gl 229B itself. 
However, two cesium lines are clearly visible in its spectrum (Oppenheimer \etal 1998).
Along with the anomalous (see Figures 2, 5, and 6) detection 
of CO in the cold atmosphere of Gl 229B alluded
to above, the atomic cesium detections may imply that its convective zone extends
below 1500 K to $\sim$1200 K and that some of its chemistry is not in equilibrium. 
In principle, grain and cloud opacities can extend the convective zone beyond the range theoretically
predicted without them (Burrows \etal 1997).
The necessity to include chemical kinetics in the convective zones will further 
complicate the theoretical treatment of brown dwarf atmospheres. 

Due to its small abundance, fluorine is not a very important element,
but its chemistry is similar to that of chlorine.  In these calculations,
most of fluorine is in HF, with much of the remainder being in the
gas--phase alkali fluorides.  Chlorine is combined in HCl, with much of the
rest at low temperatures being in the gas--phase alkali metal chlorides.

\section{Rainout}

The equilibrium abundance calculations we presented in \S 3 are fiducial
reference models in the study of the atmospheric compositions of brown 
dwarfs and EGPs, in general, and of Gl 229B, in particular.  Tables 2a and 2b and Figures 2 
through 14 summarize the results and the major trends.  However, as demonstrated
in the Gl 229B campaign and in attempts to fit its spectrum,  there is strong evidence  
that heavy metals, such as Ca, Fe, Ti, V, Si, Al, and Mg, are depleted in Gl 229B's 
atmosphere (Marley \etal 1996).   In addition, such depletions are manifest in the 
atmospheres of Jupiter and Saturn.  This is altogether to be expected, since the 
grains that condense below temperatures of $\sim$2500 K should form droplets and
rain out of the atmosphere.  This ``rainout'' will carry the condensate's elements 
to depth.  The lower boundary of the grain cloud deck should be near
where the object's temperature--pressure profile intersects the condensate's 
``Clausius--Clapyron'' lines (see Figures 2--4), but its upper boundary
is more difficult to determine.   The physical extent of the clouds
is a function of the character of convection and the meteorology of the atmosphere.
The droplet sizes are determined by the poorly--understood processes of nucleation, coagulation, and breakup.
Furthermore, the high opacity of the grains can turn an otherwise radiative
zone into a convective zone, and, thereby, influence the cloud extent and composition.
In principle, if a convective zone were well--mixed and the chemical kinetics
were suitably fast, the compositions in the convective zones would closely follow the
equilibrium abundance patterns described in \S 3.  However, the condensibles are likely to settle
and concentrate at depth. 
In short, due to the low temperatures of substellar atmospheres, 
we would expect a variety of cloud layers with a variety 
of compositions both above and below their photospheres.
Which clouds dominate is expected to be a function of the object's effective temperature, 
gravity, and metallicity.

The equilibrium calculations are a straightforward guide in determining the 
likely cloud chemicals and layers as a function of depth (temperature).  However, it is not
known where the tops of these cloud decks will reside.  The condensates that form from
temperatures of $\sim$2400 K to $\sim$1400 K ({\it e.g.}, akermanite, spinel, Fe, 
enstatite, fosterite, the titanates, MnS, high albite) have in a Gl 229B column total grammages
above a few gm cm$^{-2}$. The optical depth of a column of monodispensed
grains is approximately

\begin{equation}
\tau \sim \frac{Pf}{\rho_0ga}Q_{ext} \sim \frac{G}{\rho_0a}Q_{ext}  \, ,
\label{optical}
\end{equation}
where $P$ is the pressure at the base of the cloud layer of a given composition, $f$ is the mixing
ratio of the cloud particles, $g$ is the gravity, $Q_{ext}$ is the extinction parameter in Mie theory, $G$ is the
cloud grammage, $\rho_0$ in the density of a grain, and $a$ is the grain 
radius.  The term with $P$ in it assumes that the atmosphere is completely 
depleted in the corresponding refractory down to the cloud base.  From eq.(\ref{optical}),
we can readily calculate that the predominantly silicate and iron cloud layers below
the photosphere of Gl 229B are quite optically thick ($\ge 10^3$), the precise thickness
depending crucially upon grain radius.  Grain radii would have to exceed 1 cm
to render these clouds transparent, whereas it is expected 
that the mean $a$ will be below 100 $\mu$m\ (Lunine \etal 1989). 
Hence, these regions are expected to be more vigorously convective than would be expected
without grain opacity and the convective zone may extend to lower 
pressures and temperatures.  This phenomenon may help to explain the anomalous
detections of Cs (Oppenheimer \etal 1998) and CO (Noll, Geballe, \& Marley 1997) in Gl 229B,
particularly if disequilibrium chemistry is involved, though a detached convective zone
(Marley \etal\ 1996) may also be implicated.
 
However, the rainout of a high--temperature refractory 
will leave the upper regions of brown dwarf and EGP atmospheres 
depleted in the refractory's constituent elements,
in stoichiometric ratios.  The lower--temperature chemical species will be assembled from
the depleted elemental mixture.  As a consequence, chemical abundances will
be altered from the values assuming an Anders \& Grevesse (1989) mixture at all altitudes (\S 3). 
We attempt to explore this process with a series of artificial ``rainout'' calculations in which
we progressively sequester from the calculation the elements involved in grain formation.
From some starting temperature that defines each exercise (either 1000, 1400, or 2000 K: trial 1, 2, or 3), 
every 200 K we deplete the atmosphere (in stoichiometric ratios) of the elements found in the chemical species that
condense in that temperature interval.  In the lower 200 K interval, we perform the equilibrium calculations 
with a composition depleted in the constituent elements of the higher temperature refractories.  
This simulates in a crude way the progressive depletion and rainout of 
refractory elements from the entire atmosphere as we go down to lower 
temperatures (higher altitudes, lower pressures).   For example, for trial 3, we progressively
deplete from the atmosphere the constituents of the refractories formed at 2000 K, 1800 K, 1600 K, 1400 K, 1200 K,
1000 K, 800 K, and 600 K.  For trial 1, we do the same for 1000 K, 800 K, and 600 K, but 
with an Anders \& Grevesse mixture leave in equilibrium the layers at higher temperatures.
A major motivation for this exercise is the identification of likely condensates in the {\it upper}
atmosphere of Gl 229B, below temperatures of 1000 K, that may be implicated in 
the muting of the ``optical'' spectrum observed from $\sim$1 \mic to $\sim$0.8 \mic (Oppenheimer
\etal 1998).  The different trials represent our attempt to bracket the results.  If this
procedure consistently identifies specific low--temperature refractory 
compounds below $\sim$1000 K at useful cloud grammages then we may have
a reasonable hint at the culprits. 

Tables 3a and 3b depict the results of these trials below 1000 K.  
The temperature--pressure profile used is for a Gl 229B model with a
T$_{\rm eff}$ of 950 K and a gravity of $10^5$ cm s$^{-2}$ 
(Burrows \etal 1997).  The species listed have condensed
either in the 800--1000 K range (Table 3a) or in the 600--800 K range (Table 3b), after the corresponding depletions
have been carried out at depth by the described procedure.  Certain features are readily
apparent.  First, the grammages of the clouds formed at altitude are significantly
below the few gm cm$^{-2}$  for silicates at depth.  This is encouraging, since only
modest optical depths are needed to explain the shorter--wavelength Gl 229B data.  
From eq.(\ref{optical}), a grammage of only $10^{-5}$  gm cm$^{-2}$ with a particle radius of
0.1 \mic provides close to unit optical depth.  (Note that small particles are expected in radiative zones.)
Second, the ``abundant'' low--temperature refractories are 
predominantly the chlorides and the sulfides.  In particular,
KCl, NaCl, NaF, and SiS$_2$ stand out, with SiS$_2$ dominating for trial 3.  Without depletion, SiS$_2$
would not form in abundance, but progressive depletion starting at the highest temperatures
is needed for it to become important.  If SiS$_2$ forms, it does so around 743 K,
whereas NaF forms near 920 K, NaCl forms near 800 K, and KCl forms near 740 K.
Be that as it may, the sodium and potassium salts emerge
as cloud candidates in Gl 229B's upper atmosphere, with grammages that may range from $\sim 10^{-4}$
to  $\sim 4\times 10^{-6}$ gm cm$^{-2}$.  In trial 3, though the KCl cloud is thin, the SiS$_2$ is thick.
These calculations suggest that thin clouds of non--silicate, low--temperature refractories
can and do exist in the upper atmosphere of Gl 229B.  Furthermore, they  may also reside in the 
atmospheres of brown dwarfs with T$_{\rm eff}$s below about 1400 K (?). 
Hence, it is expected that many cloud layers with different compositions 
are formed both above and below the photospheres
of brown dwarfs and EGPs.  At times, from the lowest effective temperatures up, clouds of 
either NH$_3$ (T$_{\rm eff}$ $\le$ 250 K), H$_2$O (T$_{\rm eff}$ $\le$ 400 K), chlorides,
sulfides, iron, or silicates should be found in these 
exotic substellar and late M dwarf atmospheres.

\section{Discussion and Conclusions}

We have explored in detail the abundance profiles and 
compositions expected in the atmospheres
of brown dwarfs and EGPs.  In chemical equilibrium, the major reservoirs   
of the dominant elements shift with pressure and, most importantly,
with temperature in ways reflected in Figures 2 through 14 and in Tables
2a and 2b.  Unlike in any other stellar context, chemistry and molecules
assume a central role in determining the character of the atmospheres 
of substellar objects.  As the temperature decreases with increasing 
height in such an atmosphere, molecules
not encountered in M dwarfs or in standard stellar 
atmospheres form.   At the higher temperatures,
the standard refractories, such as the silicates, spinel, and iron, 
condense out into grain clouds which by their large opacity 
lower the T$_{\rm eff}$ and luminosity of the
main sequence edge (Burrows \etal 1993) and alter 
in detectable ways the spectra of objects
around the transition mass (Jones \& Tsuji 1997).  As T$_{\rm eff}$ decreases 
below that at the stellar edge, the classical refractories
are buried progressively deeper below the photosphere and less refractory
condensates and gas--phase molecules come to dominate 
(Marley \etal 1996; Burrows \etal 1997).  

Below temperatures of
$\sim$1500 K, our calculations demonstrate and confirm that 
the alkali metals, which are not as refractory
as Fe, Al, Ca, Ti, V, and Mg, emerge as important 
atmospheric and spectral constituents.  At still lower temperatures,
chlorides and sulfides appear, some of which will condense in the cooler
upper atmosphere and form clouds that will affect emergent spectra and
albedos.  Cloud decks of many different compositions at many different
temperature levels are expected, depending upon T$_{\rm eff}$ (and weakly
upon gravity).  We expect that clouds of chlorides and sulfides (not silicates), 
at temperature levels below $\sim$1000 K, are responsible
for the steeper slope observed in the spectrum of Gl 229B 
at the shorter wavelengths (Oppenheimer \etal 1998).
At slightly higher temperatures, MnS, ZnS, NaAlSi$_3$O$_8$, 
KAlSi$_3$O$_8$, V$_2$O$_3$, Mg$_3$P$_2$O$_8$,
and MgTi$_2$O$_5$ may play a role, but only if 
their constituents are not scavenged into more
refractory compounds and rained out deeper down.

As T$_{\rm eff}$ decreases (either as a given mass cools or, for a given age,
as we study objects with lower masses), the major atmospheric constituents of  
brown dwarfs and EGPs change. This change is reflected in  
which spectral features are most prominent  
and in the albedos of substellar objects near their primaries.
Hence, specific mixes of atoms, molecules, and clouds can serve
as approximate T$_{\rm eff}$ and temperature indicators and a 
composition scale can be established.  In order to do this definitively, synthetic
spectra with the atmospheres we have calculated are required.
However, the composition trends we have identified are suitably dramatic that
reasonable molecular indicators of spectral type can be suggested. 
A workable sequence might be: TiO disappears (at 2300--2000 K), 
refractory silicates and Fe(c) appear
(at 2300--2000 K), Mg$_2$SiO$_4$ appears (at 1900 K), VO(g) disappears (at 1700--1900 K), 
MgSiO$_3$ appears (at 1700 K), silicates rainout (at $\sim$1400 K (?)), 
CrH disappears ($\sim$1400 K), Li $\rightarrow$ LiCl 
($\le$1400 K), CO $\rightarrow$ CH$_4$ (1200--1500 K), 
(Rb,Cs,K) $\rightarrow$ chlorides ($\le$ 1200 K), PH$_3$ $\rightarrow$ (P$_4$O$_6$, Mg$_3$P$_2$O$_8$)
($\le$ 1000 K), formation of NaF, NaCl, and KCl clouds and 
various sulfide clouds ($\sim$700--1100 K),
N$_2$ $\rightarrow$ NH$_3$(g) ($\sim$700 K), 
H$_2$O(g) $\rightarrow$ H$_2$O(c) ($\sim$350 K), and 
NH$_3$ $\rightarrow$ (NH$_3$(c), NH$_4$SH(c)) ($\sim$200 K).  

Disequilibrium chemistry and convection
and differences in the spectroscopic strengths of the various
indicators will no doubt partially alter the T$_{\rm eff}$ order of this spectral scale.
Note that objects with T$_{\rm eff}$s higher than the temperatures quoted above
may nevertheless, because of the lower temperatures that can be achieved
at lower optical depths, manifest lower--temperature compounds and/or transitions
in their atmospheres and spectra.  Gl 229B with a T$_{\rm eff}$ of $\sim$950 K
is a case in point, if lower--temperature chloride and sulfide refractories 
are to be invoked to explain its spectrum.   How much higher the appropriate T$_{\rm eff}$
should be for a given chemical transition to be fully manifest will depend upon consistent
atmosphere models.  Clearly, the above temperatures for the appearance or disappearance
of species should be used with great caution when estimating T$_{\rm eff}$.
Very crudely, the ``L'' spectral type suggested by Kirkpatrick \etal (1998) would 
correspond to T$_{\rm eff}$s between about $\sim$1500 K and $\sim$2200 K.
None but the youngest and most massive brown dwarfs and 
only the very youngest EGPs could have this proposed spectral designation.  Most brown dwarfs and EGPs
will be of an even later spectral type, yet to be coined, a spectral type that would include Gl 229B. 

In the future, more consistent atmospheres and a detailed spectral 
sequence need to be calculated.  Furthermore, we need to expand our
inventory of possible compounds and explore the 
effect of metallicity.  Importantly, credible cloud models,
incorporating realistic grain size distributions and their altitude dependence, 
will be required to explain the spectral and photometric data on brown dwarfs
and extrasolar giant planets that has already 
been obtained and that can be anticipated in the
near future.  Given the ambiguities in cloud physics 
and theoretical meteorology, new observations
will be crucial to educate and guide theory concerning the true nature
of substellar atmospheres.  Nevertheless, it is clear that brown dwarfs 
and giant planets occupy an exciting new realm of science midway between astronomy
and planetary studies in which new insights and for which new tools
must be developed.  The theoretical study of substellar objects with T$_{\rm eff}$s
between those of Jupiter and the stars has only just now begun in earnest.

\acknowledgements

We thank Mark Marley, Bill Hubbard, Jonathan Lunine, 
David Sudarsky, Didier Saumon, Shri Kulkarni, Ben Oppenheimer, Jim Liebert,
Davy Kirkpatrick, France Allard, Gilles Chabrier, and 
Tristan Guillot for a variety of useful contributions.
This work was supported under NASA grants NAG5-7499, NAG5-7073, and NAG5-2817.

\clearpage

\figcaption{{
Plot of the temperature--pressure profiles for models with masses
of 5, 10, 20 and 40 M$_{\rm J}$ at ages of 0.1, 1, and 10 Gyr shown as dash--dotted,
dashed, and solid curves, respectively.   
The positions of the photospheres for each model are shown as solid circles.
The four horizontal lines connect the positions 
of the photospheres for models with the same mass. 
(The lowest mass (5 M$_{\rm J}$) line--of--photospheres is at the top.)
The masses in M$_{\rm J}$s label the corresponding profiles.
At 0.1 Gyr, there are only two models and labels, 
one for 5 M$_{\rm J}$ and one for 10 M$_{\rm J}$.  Note that at high pressures the model 
for 5 M$_{\rm J}$ at 1 Gyr overlaps with that for 40 M$_{\rm J}$ at 10 Gyr.
}}

\figcaption{{
Condensation or chemical transformation boundaries for 
major species in temperature--pressure space. Superposed on the figure are
atmospheric profiles, shown as solid curves, from Figure 1 for a
number of giant planet and brown dwarf models with an age of 1 Gyr.
The models of Burrows \etal (1997) were employed.
Included are a present--day Jupiter profile and 1 Gyr adiabats for
M dwarf models with  masses of 0.08, 0.09 and 0.115 M$_{\odot}$
($1 {\rm M}_{\odot} = 1047 {\rm M}_{\rm J}$).  The black-filled circles indicate the location of the
photosphere.  The region to the left of a circle spans the atmosphere above the 
photosphere.  Note that for the red dwarfs we have
extrapolated temperatures somewhat above the height of the photosphere assuming
an adiabatic law.  In reality, this is in the radiative regime and the profile pressures
at a given temperature should be lower, as are those for the substellar models.
Chemical boundary curves indicate when a condensed species forms
or when a ratio of abundances is unity.  The condensates are written in braces
``\{ \}'' to distinguish them from gas--phase species.  A condensate forms on
and immediately to the left of each indicated curve, and in some cases may
disappear into another phase at lower temperatures, but is not shown.  
The curve labeled by a ratio, such as CH$_4$/CO, indicates where these two
species have equal abundances, with the ratio greater than unity to the left.
See the text for further explanations.
}}

\figcaption{{
The same as Figure 2, except important silicates, spinel,
and phosphorus compounds are plotted.  As in Figure 2,  
the ratios or abundances increase to the left. See text for details.
}}

\figcaption{{
The same as Figure 2, but showing the loci of points where the
partial pressures of the gas-phase alkali metal chlorides are equal to those
of the corresponding metal vapors.  The chlorides exist on the 
low--temperature side of the metal/chloride boundaries.  With the 
exception of lithium, which has some characteristics similar to an
alkaline--earth element, the formation of the chloride occurs at 
temperatures that increase with increasing molecular weight.
}}

\figcaption{{
Plots of the abundance profiles for several of the most abundant chemical
species, H$_2$O, CH$_4$, CO, NH$_3$ and N$_2$, for brown dwarf 
models with an effective temperature of 950 K,
but with the gravities of 300, 1000 and 3000 m s$^{-2}$.  The temperature of
950 K is chosen because it is near that of the brown dwarf Gl 229B.  
Temperature--pressure profiles for the underlying dwarf models are implicit
and are from Burrows \etal (1997).  Each curve is a plot
of $\log_{10}$ of the number fraction of a particular species out of all
species, including any condensates, as a function of level temperature. 
An increase in temperature generally corresponds to an increase in
pressure and depth.  Each set of curves is labeled by the chemical species
H$_2$O, CH$_4$, CO, NH$_3$ and N$_2$, and the numbers 300, 1000 and 3000
correspond to each model's gravity.  For example, the set of curves labeled
CO and 300, 1000, and 3000 show the profiles for CO for the three models.
The two most abundant species, H$_2$ and He, would be above the top of the
figure and are not plotted.  For clarity, the profile for CO is shown as a
dashed curve and the label of 1000 is omitted from the plots of H$_2$O and
N$_2$.  The vertical line at 950 K shows the position of the photosphere.
The condensates are identified here and in
the following figures by being enclosed in braces, \{\}.
}}

\figcaption{{
The same as Figure 5 for the same five species, but
for five models with the effective temperatures of 800, 850, 900, 950
and 1000 K, for a fixed gravity of 1000 m s$^{-2}$.  Each set of curves is
labeled by the species, with 800 and 1000 indicating the coolest and hottest
models in the sequence, with the other models lying in between.  The five
photospheric temperatures are shown as short tick marks at the bottom of the
figure.
}}

\figcaption{{
The abundances of the
gas--phase species H$_2$S, together with the condensates Fe, FeS, MgSiO$_3$
and Mg$_2$SiO$_4$, are plotted for the  effective temperature of 950 K,
but with the gravities of 300, 1000 and 3000 m s$^{-2}$, as in Figure 5.  As the abundances of 
H$_2$S and FeS are independent of the model, the numbers
labeling the gravities are omitted.
}}

\figcaption{{
The same five species are plotted as in Figure 7, but for 
the five models with fixed gravity and varying effective temperature used in
Figure 6.
}}

\figcaption{{
The abundances for NaAlSi$_3$O$_8$, KAlSi$_3$O$_8$, 
MgAl$_2$O$_4$, Mg$_3$P$_2$O$_8$, P$_4$O$_6$,
PH$_3$, MnS, CaMgSi$_2$O$_6$, Ca$_2$MgSi$_2$O$_7$, and Ca$_2$Al$_2$SiO$_7$,
for the gravities and effective temperature of Figure 5.
For clarity, different line thicknesses are used.  
This figure includes some of the refractory silicates not
plotted in the previous figures.  From right to left,
corresponding to decreasing temperature, the first condensate shown is
Ca$_2$Al$_2$SiO$_7$, identified by the thick curves.  Only the 3000~m s$^{-2}$
gravity model is labeled.  The thinner curves labeled ``300'' and
``3000'' at the top are for MgAl$_2$O$_4$, and continue on to lower
temperatures.  The next set of thinner curves on either side of 2000~K
refer to the appearance of Ca$_2$MgSi$_2$O$_7$, for which the 300 and 1000
models are labeled immediately to the left.  The three following curves
which rise to above $-5.5$ in the ordinate below 1900~K refer to the
appearance of CaMgSi$_2$O$_6$, with the accompanying curves of the previous
species indicating its disappearance.  The remaining set of curves at lower
temperatures are clearly identified by at least one of the models.
}}

\figcaption{{
The same ten species shown in Figure 9, with the same
line thickness, are plotted for models with effective temperatures of 800, 850, 900, 950
and 1000 K, for a fixed gravity of 1000 m s$^{-2}$, as in Figure 6.
The first condensate to be plotted to the right 
is Ca$_2$Al$_2$SiO$_7$, as in Figure 9, and 
is for the 800 K model.
}}

\figcaption{{
The titanium and vanadium condensates, MgTi$_2$O$_5$,
Ti$_4$O$_7$, Ti$_2$O$_3$, CaTiO$_3$, V$_2$O$_3$ and VO, together with
their most important gas--phase species, TiO and VO, are plotted for the
three different gravity models depicted in Figure 5 and for an 
effective temperature of 950 K.  Note that with
decreasing temperature, the highest temperature condensate, Ti$_3$O$_5$,
changes into Ti$_2$O$_3$ below 1900~K for the two higher gravity models,
but for the lowest 300 m s$^{-2}$ model Ti$_3$O$_5$ temporarily changes
into CaTiO$_3$ above 2000~K in a range of about 100~K, before changing
back into the high temperature condensate.  As with the other models, this
then changes into Ti$_2$O$_3$ below 1900~K.
}}

\figcaption{{
The abundances of the same titanium and vanadium 
species depicted in Figure 11 are plotted  for models with 
effective temperatures of 800, 850, 900, 950
and 1000 K, for a fixed gravity of 1000 m s$^{-2}$, as in Figure 6. 
Note that in these models CaTiO$_3$ does not form.
}}

\figcaption{{
The abundances of lithium, rubidium,
cesium and their most abundant compounds for an effective temperature of
950 K and the three gravities employed in Figure 5. At the lowest temperatures
plotted, the condensate RbCl forms, with the 300 m s$^{-2}$ model being
labeled.
}}

\figcaption{{
The abundances of the same species depicted in Figure 13 are plotted for
models with effective temperatures of 800, 900, 
and 1000 K, for a fixed gravity of 1000 m s$^{-2}$, as in Figure 6.
We show a restricted set of curves because the results are only
weakly dependent on effective temperature.
}}

\clearpage

\begin{deluxetable}{clcl}
\tablewidth{15cm}
\tablenum{1}
\tablecaption{Anders and Grevesse (1989) solar abundances \tablenotemark{\dag}}
\tablehead{
\colhead{Element} & \colhead{Abundance} & \colhead{Element} & \colhead{Abundance}}
\startdata
H   &  $9.10\times 10^{-1}$  &  Ni  &  $1.61\times 10^{-6}$     \nl
He  &  $8.87\times 10^{-2}$  &  Cr  &  $4.40\times 10^{-7}$     \nl
O   &  $7.76\times 10^{-4}$  &  P   &  $3.39\times 10^{-7}$     \nl
C   &  $3.29\times 10^{-4}$  &  Mn  &  $3.11\times 10^{-7}$     \nl
Ne  &  $1.12\times 10^{-4}$  &  Cl  &  $1.71\times 10^{-7}$     \nl
N   &  $1.02\times 10^{-4}$  &  K   &  $1.23\times 10^{-7}$     \nl
Mg  &  $3.49\times 10^{-5}$  &  Ti  &  $7.83\times 10^{-8}$     \nl
Si  &  $3.26\times 10^{-5}$  &  Co  &  $7.34\times 10^{-8}$     \nl
Fe  &  $2.94\times 10^{-5}$  &  F   &  $2.75\times 10^{-8}$     \nl
S   &  $1.68\times 10^{-5}$  &  V   &  $9.56\times 10^{-9}$     \nl
Ar  &  $3.29\times 10^{-6}$  &  Li  &  $1.86\times 10^{-9}$     \nl
Al  &  $2.77\times 10^{-6}$  &  Rb  &  $2.31\times 10^{-10}$    \nl
Ca  &  $1.99\times 10^{-6}$  &  Cs  &  $1.21\times 10^{-11}$    \nl
Na  &  $1.87\times 10^{-6}$

\tablenotetext{\dag}{The elements are given as fractions by number in order of
decreasing abundance.}
\enddata
\end{deluxetable}

\clearpage

\begin{deluxetable}{cl}
\tablewidth{15cm}
\tablenum{2a}
\tablecaption{The most important species associated with each element \tablenotemark{\dag}}
\tablehead{
\colhead{Element} & \colhead{Major Chemical Species}}
\startdata
H  & H$_2$                                                    \nl
He & He                                                       \nl
Li & Li LiCl LiF \undertext{LiAlO$_2$}                        \nl
C  & CO CH$_4$                                           \nl
N  & N$_2$ NH$_3$ \undertext{NH$_3$}                           \nl
O  & H$_2$O CO \undertext{H$_2$O}                              \nl
F  & HF LiF NaF KF RbF CsF \undertext{Na$_3$AlF$_6$} \undertext{LiF} \nl
Ne & Ne                                                  \nl
Na & Na NaCl \undertext{NaCl} \undertext{NaAlSi$_3$O$_8$} \undertext{Na$_3$AlF$_6$} \nl
Mg & Mg MgH \undertext{MgSiO$_3$} \undertext{Mg$_2$SiO$_4$}
\undertext{MgAl$_2$O$_4$} \undertext{CaMgSi$_2$O$_6$} \nl
   &  \undertext{Ca$_2$MgSi$_2$O$_7$} \undertext{Mg$_3$P$_2$O$_8$} \undertext{MgTi$_2$O$_5$} \nl
Al & Al AlH AlOH Al$_2$O \undertext{Al$_2$O$_3$}
\undertext{MgAl$_2$O$_4$} \undertext{Ca$_2$Al$_2$SiO$_7$}
\undertext{NaAlSi$_3$O$_8$} \nl
   &  \undertext{KAlSi$_3$O$_8$} \undertext{Na$_3$AlF$_6$} \undertext{LiAlO$_2$} \nl
Si & SiO \undertext{MgSiO$_3$} \undertext{Mg$_2$SiO$_4$}
\undertext{Ca$_2$Al$_2$SiO$_7$} \undertext{Ca$_2$SiO$_4$}
\undertext{CaMgSi$_2$O$_6$} \nl
   &  \undertext{Ca$_2$MgSi$_2$O$_7$} \undertext{NaAlSi$_3$O$_8$}
\undertext{KAlSi$_3$O$_8$} \undertext{MnSiO$_3$} \nl
P  & PH$_3$ P$_4$O$_6$ PN \undertext{Mg$_3$P$_2$O$_8$}         \nl
S  & H$_2$S SH SiS \undertext{MnS} \undertext{FeS} \undertext{NH$_4$SH}     \nl
\tablenotetext{\dag}{The underlined species are condensates and
the gas--phase species are given first.  Otherwise, the order of the
species is generally unimportant, though some
condensates of less importance are given last. }
\enddata
\end{deluxetable}

\clearpage

\begin{deluxetable}{cl}
\tablewidth{15cm}
\tablenum{2b}
\tablecaption{Continuation of the listing of the most important species associated with each element \tablenotemark{\dag}}
\tablehead{
\colhead{Element} & \colhead{Major Chemical Species}}
\startdata
Cl & HCl LiCl NaCl KCl RbCl KCl AlCl MgCl CaCl \undertext{NaCl} \undertext{RbCl} \undertext{CsCl} \nl
Ar & Ar                                                  \nl
K  & K KCl \undertext{KAlSi$_3$O$_8$}                            \nl
Ca & Ca CaH CaOH \undertext{Ca$_2$Al$_2$SiO$_7$} \undertext{Ca$_2$SiO$_4$}
\undertext{Ca$_2$MgSi$_2$O$_7$} \undertext{CaMgSi$_2$O$_6$} \nl
Ti & Ti TiO \undertext{Ti$_3$O$_5$} \undertext{Ti$_2$O$_3$}
\undertext{Ti$_4$O$_7$} \undertext{MgTi$_2$O$_5$} \undertext{CaTiO$_3$} \nl
V  & V VO \undertext{VO} \undertext{V$_2$O$_3$} \undertext{VN}               \nl
Cr & Cr CrH \undertext{Cr} \undertext{Cr$_2$O$_3$}                       \nl
Mn & Mn MnH \undertext{MnS} \undertext{MnSiO$_3$}                        \nl
Fe & Fe FeH \undertext{Fe} \undertext{FeS}                               \nl
Co & Co \undertext{Co}                                         \nl
Ni & Ni NiH \undertext{Ni}                                         \nl
Rb & Rb RbCl RbF \undertext{RbCl}                                    \nl
Cs & Cs CsCl CsF \undertext{CsCl}                                        \nl
\tablenotetext{\dag}{The underlined species are condensates and
the gas--phase species are given first.  Otherwise, the order of the
species is generally unimportant, though some
condensates of less importance are given last. }
\enddata
\end{deluxetable}

\clearpage

\begin{deluxetable}{lccclll}
\tablewidth{15cm}
\tablenum{3a}
\tablecaption{Inferred cloud grammages (g cm$^{-2}$) for species that condense between 1000 and 800 K
for the three rainout trials\tablenotemark{\dag}}
\tablehead{
\colhead{Condensate} & \colhead{A1} & \colhead{A2} & \colhead{A3} &
\colhead{Grammage 1} & \colhead{Grammage 2} & \colhead{Grammage 3}}
\startdata

MnO          & 927 &     &     & $2.09\times10^{-10}$ \nl
FeS          & 875 &     &     & $6.90\times10^{-11}$ \nl
NaF          &     & 924 &     &                      & \undertext{$2.01\times10$}$^{-5}$  \nl
Ni$_3$S$_4$  &     & 922 &     &                      & $1.26\times10^{-9}$  \nl
Fe$_3$O$_4$  &     & 921 &     &                      & $2.14\times10^{-8}$  \nl
Co$_3$O$_4$  &     & 836 &     &                      & $1.17\times10^{-8}$  \nl
NaCl         &     & 825 &     &                      & \undertext{$6.74\times10$}$^{-5}$  \nl
Li$_2$O      &     & 800 &     &                      & $5.53\times10^{-9}$  \nl
CaCl$_2$     &     &     & 863 &                      &                      & $2.45\times10^{-8}$  \nl
Fe$_2$O$_3$  &     &     & 862 &                      &                      & $1.18\times10^{-7}$  \nl
NiS$_2$      &     &     & 835 &                      &                      & $1.15\times10^{-7}$  \nl

\tablenotetext{\dag}{A1, A2 and A3 are the temperatures in Kelvin when the condensate first appears for
the first, second, and third rainout trials, respectively.}
\enddata
\end{deluxetable}

\clearpage

\begin{deluxetable}{lccclll}
\tablewidth{15cm}
\tablenum{3b}
\tablecaption{Inferred cloud grammages (g cm$^{-2}$) for species that condense between 800 and 600 K
for the three rainout trials\tablenotemark{\dag}}
\tablehead{
\colhead{Condensate} & \colhead{A1} & \colhead{A2} & \colhead{A3} &
\colhead{Grammage 1} & \colhead{Grammage 2} & \colhead{Grammage 3}}
\startdata

NaCl         & 776 &     & 701 & \undertext{$1.04\times10$}$^{-5}$  &                      & $2.44\times10^{-7}$  \nl
KCl          & 730 & 740 & 728 & \undertext{$4.62\times10$}$^{-6}$  &
\undertext{$3.68\times10$}$^{-5}$  &\undertext{$4.20\times10$}$^{-6}$  \nl
LiF          & 700 &     &     & $2.68\times10^{-7}$  \nl
RbCl         & 613 &     & 613 & $4.94\times10^{-8}$  &                      & $4.94\times10^{-8}$  \nl
Mn           &     & 768 & 768 &                      & $2.16\times10^{-9}$  & $2.12\times10^{-9}$  \nl
LiOH         &     & 745 &     &                      & $3.15\times10^{-7}$  \nl
Fe$_2$O$_3$  &     & 726 &     &                      & $1.72\times10^{-10}$ \nl
MgF$_2$      &     & 636 &     &                      & $5.77\times10^{-10}$ \nl
KF           &     & 605 &     &                      & $5.03\times10^{-9}$  \nl
SiS$_2$      &     &     & 743 &                      &                      & \undertext{$4.13\times10$}$^{-3}$  \nl
Co$_3$O$_4$  &     &     & 684 &                      &                      & $5.52\times10^{-9}$  \nl
LiCl         &     &     & 638 &                      &                      & $2.63\times10^{-7}$  \nl
CaF$_2$      &     &     & 627 &                      &                      & $1.77\times10^{-10}$ \nl
MgCl$_2$     &     &     & 615 &                      &                      & $8.85\times10^{-10}$ \nl

\tablenotetext{\dag}{A1, A2 and A3 are the temperatures in Kelvin when the condensate first appears for the
first, second, and third rainout trials, respectively.}
\enddata
\end{deluxetable}

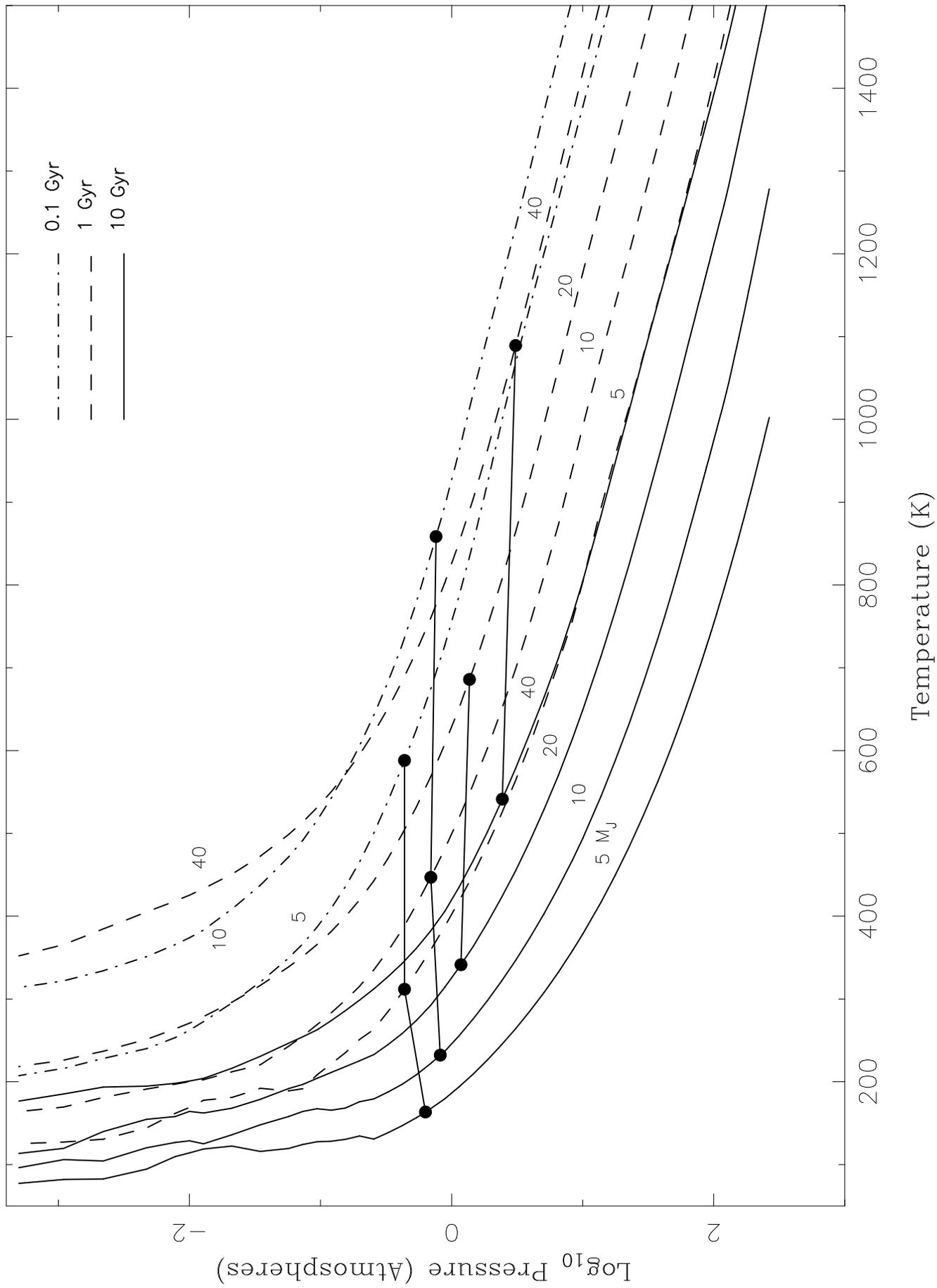
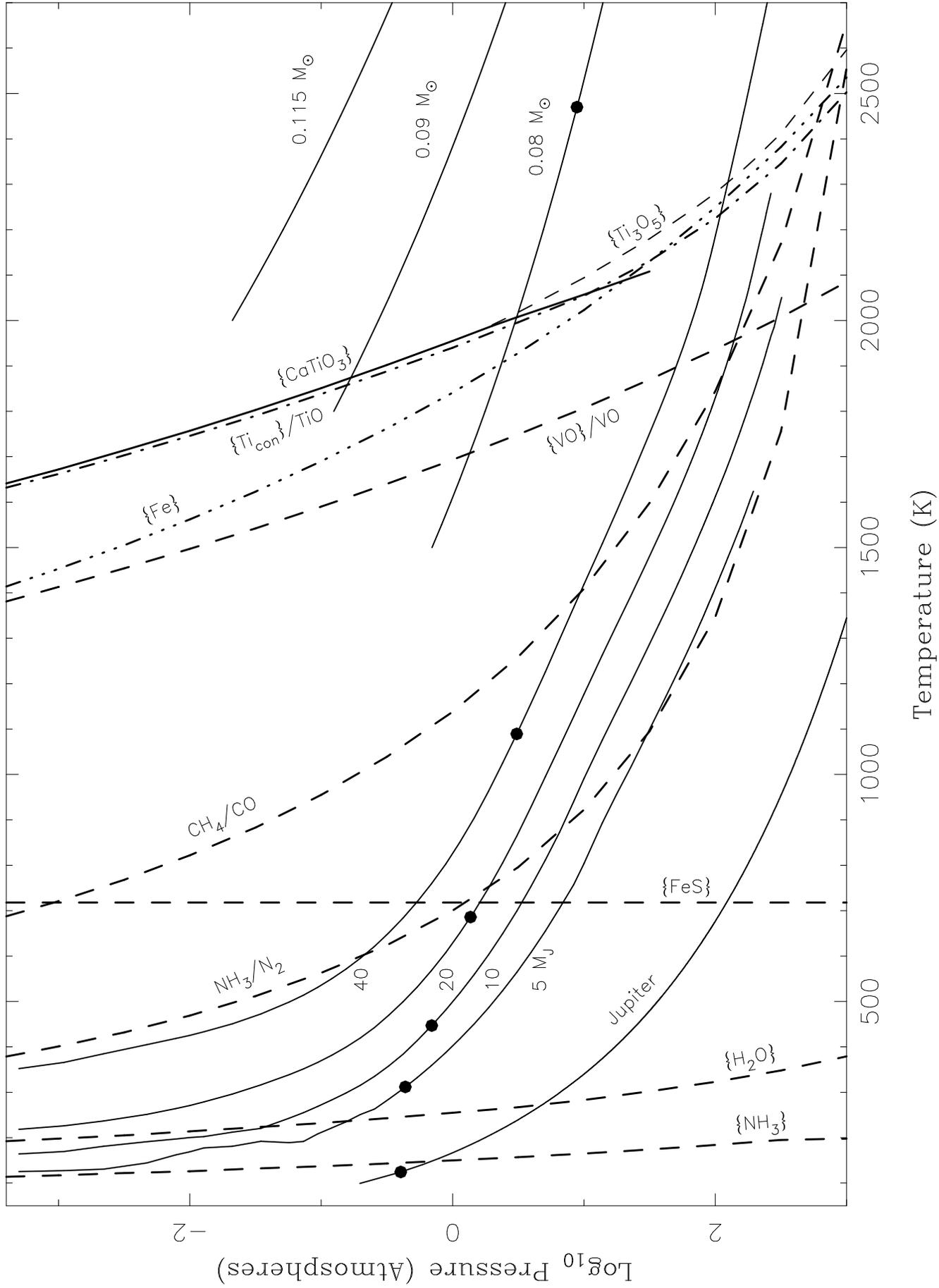
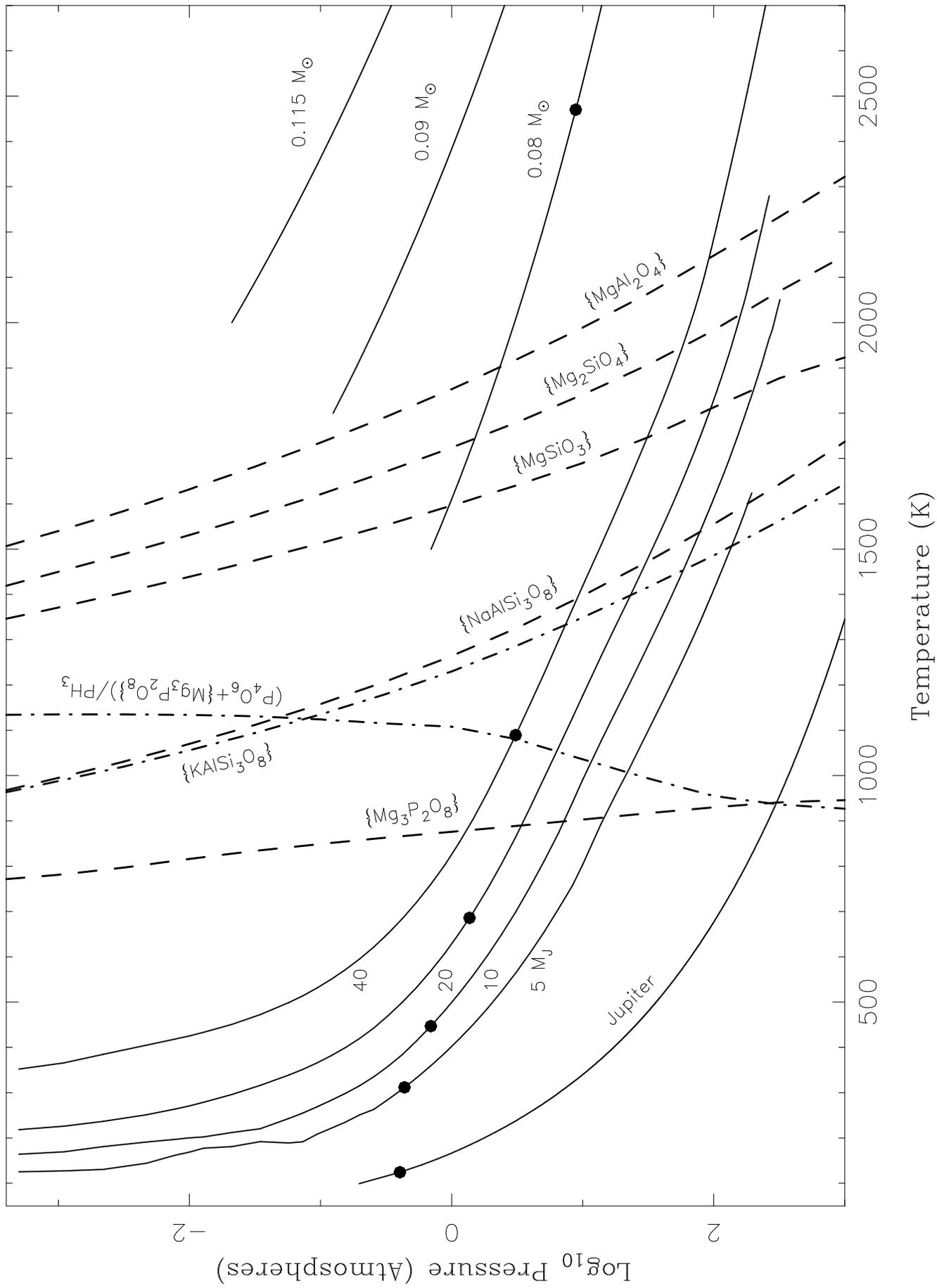
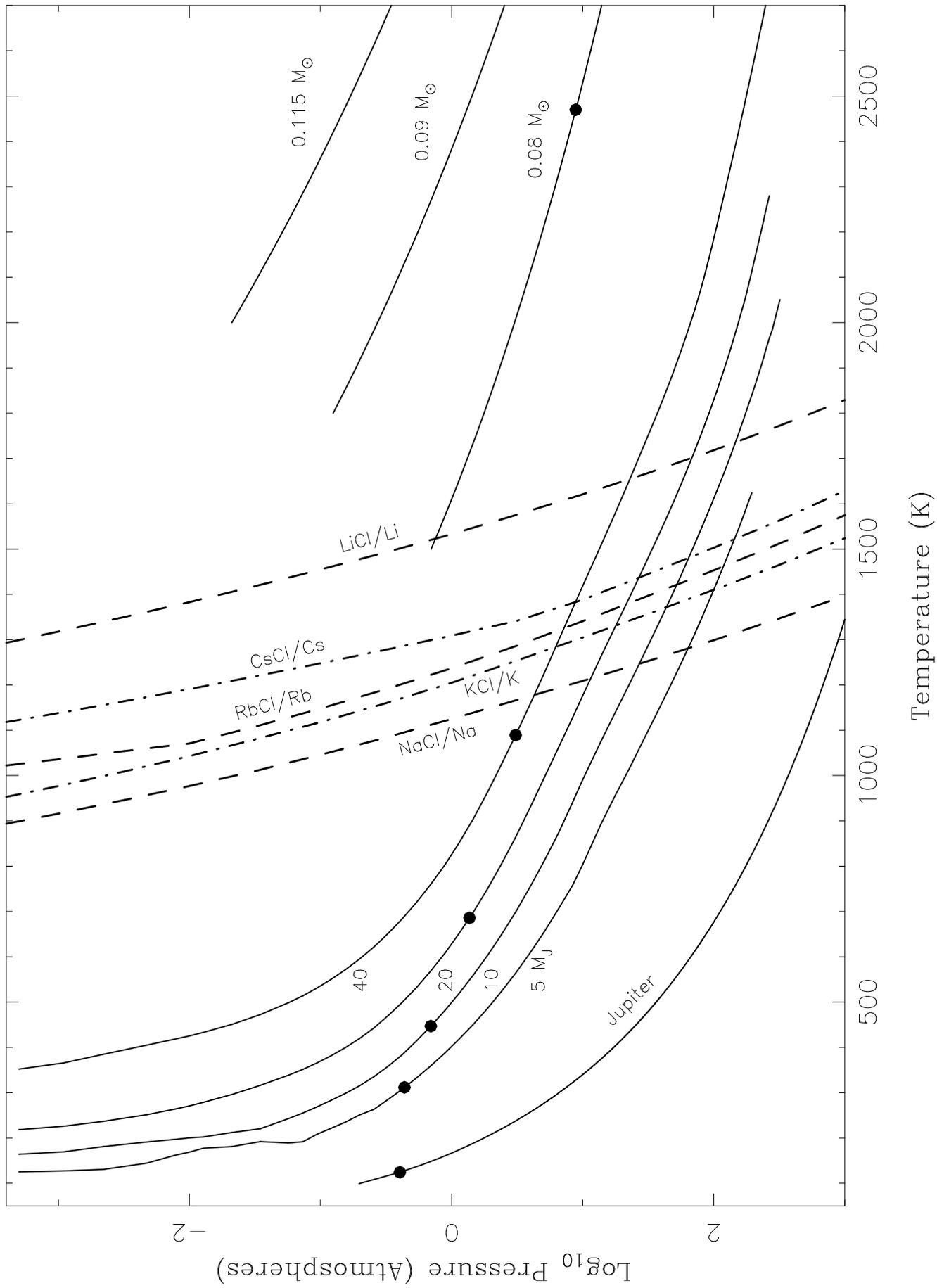
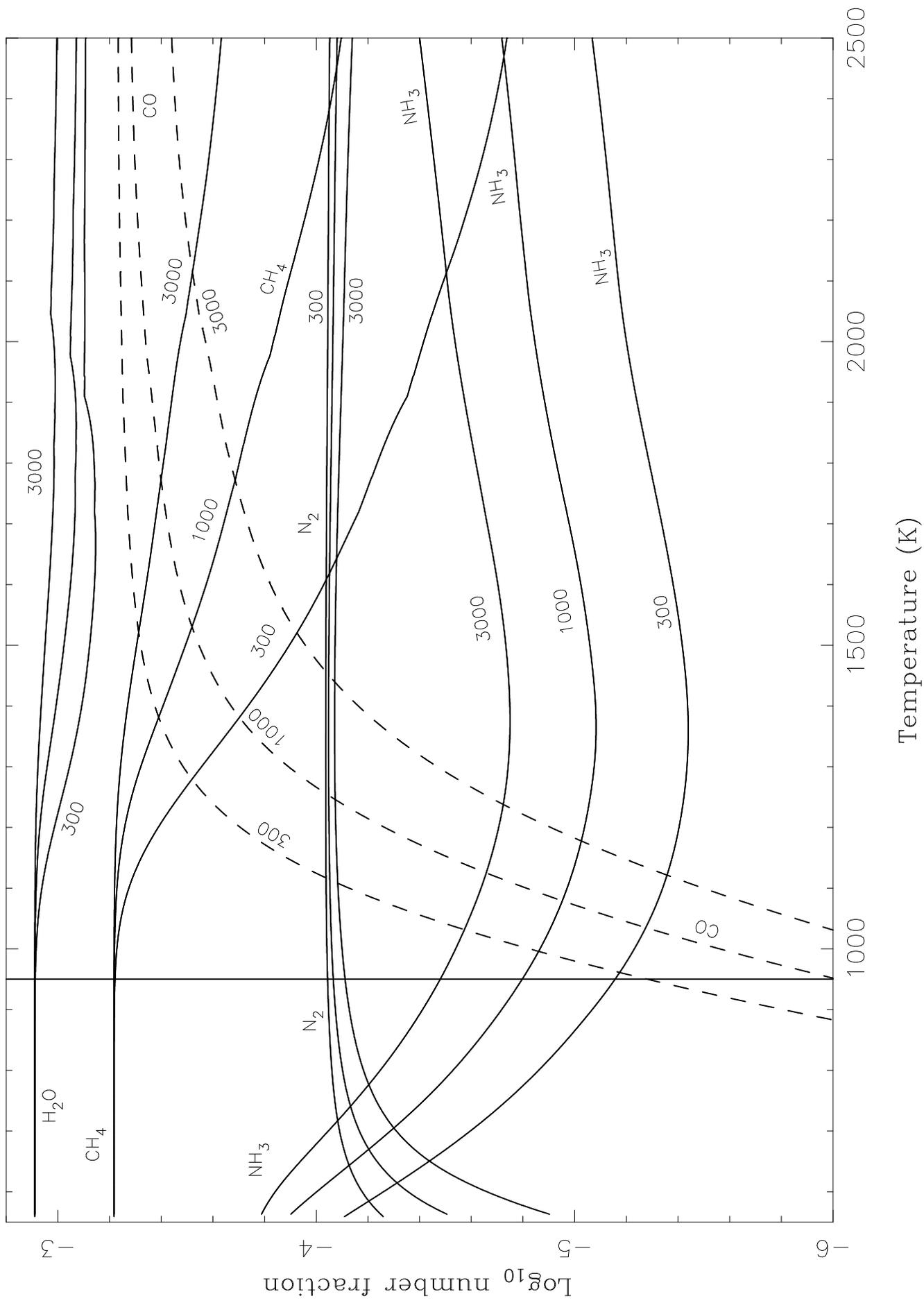
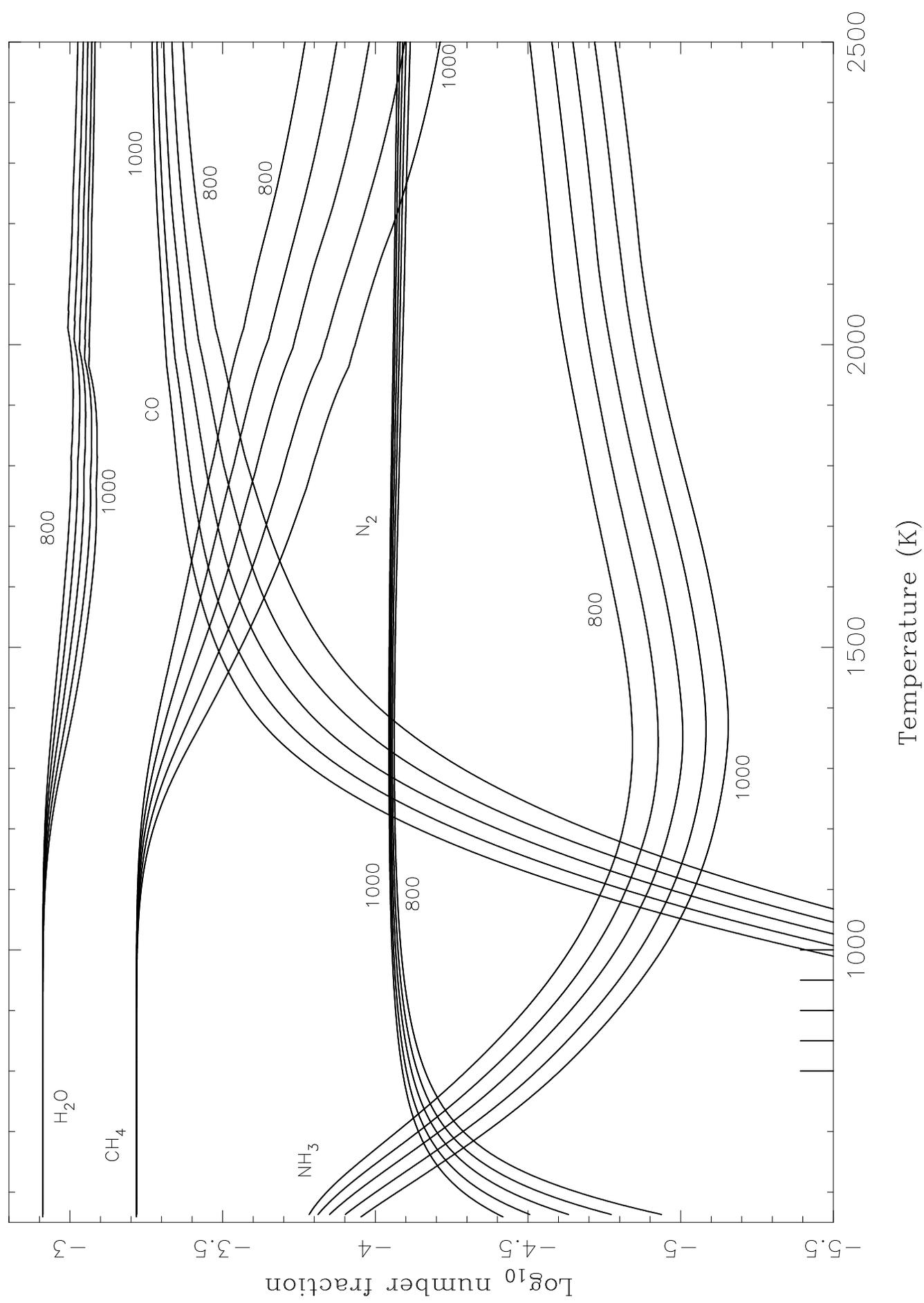
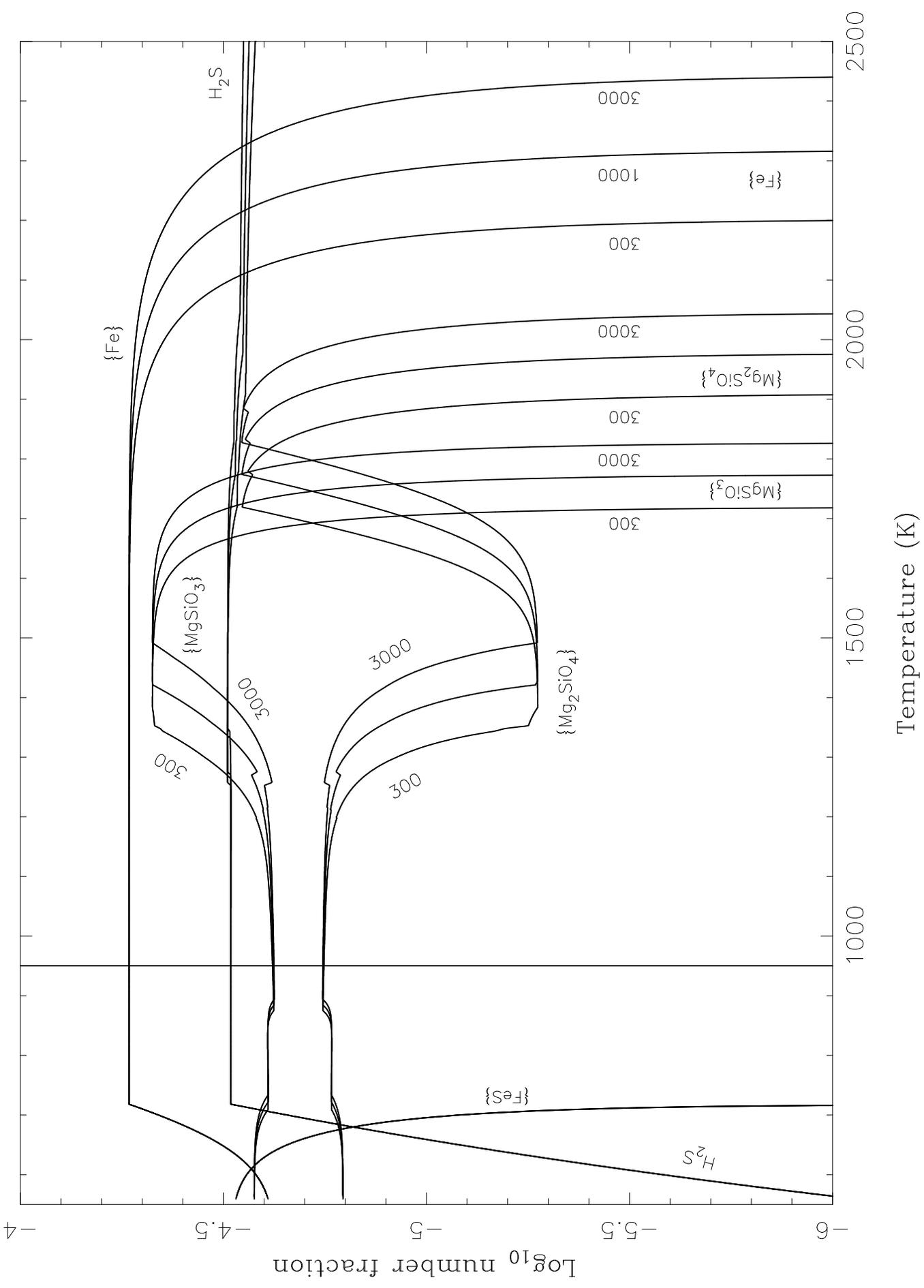
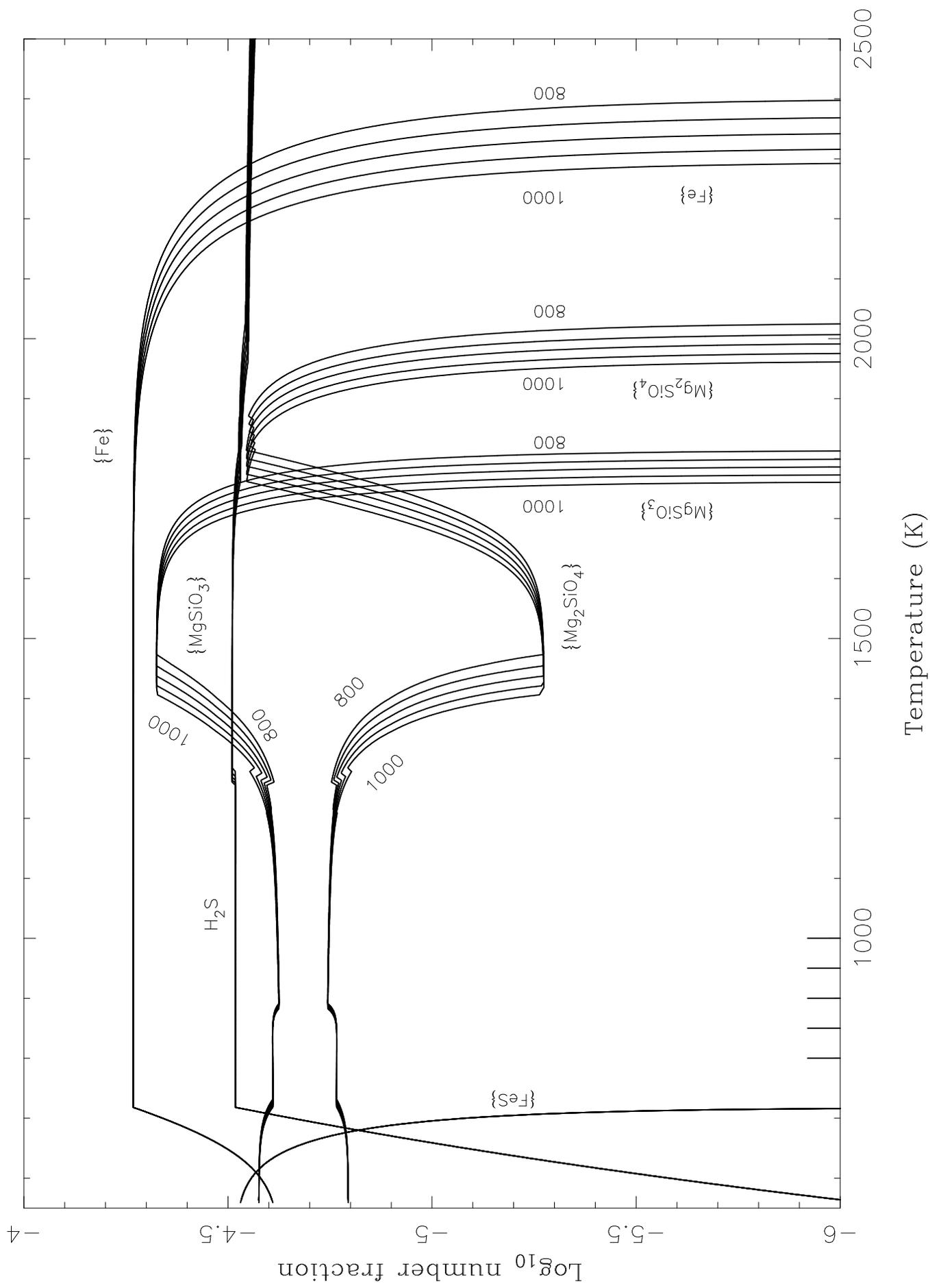
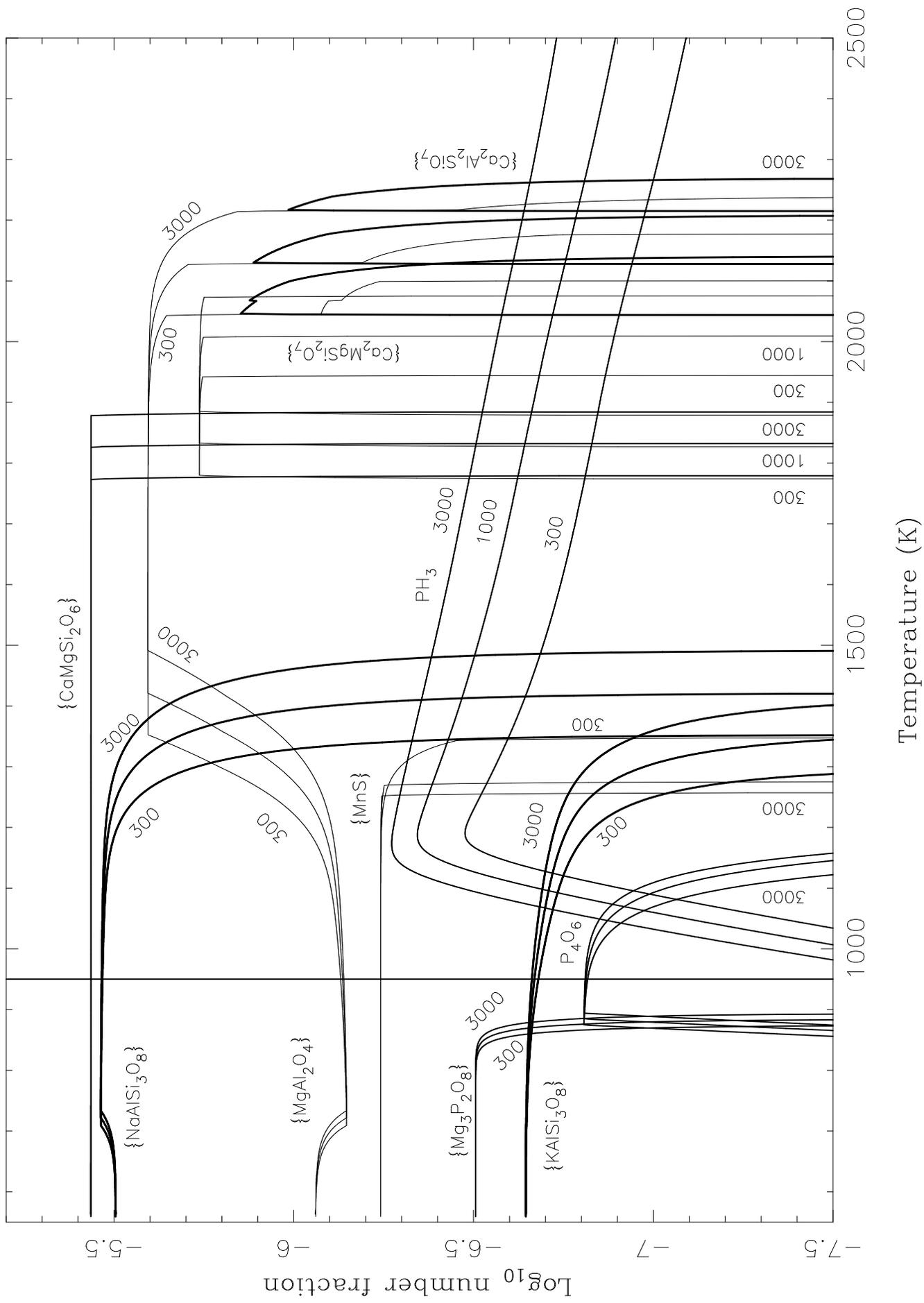
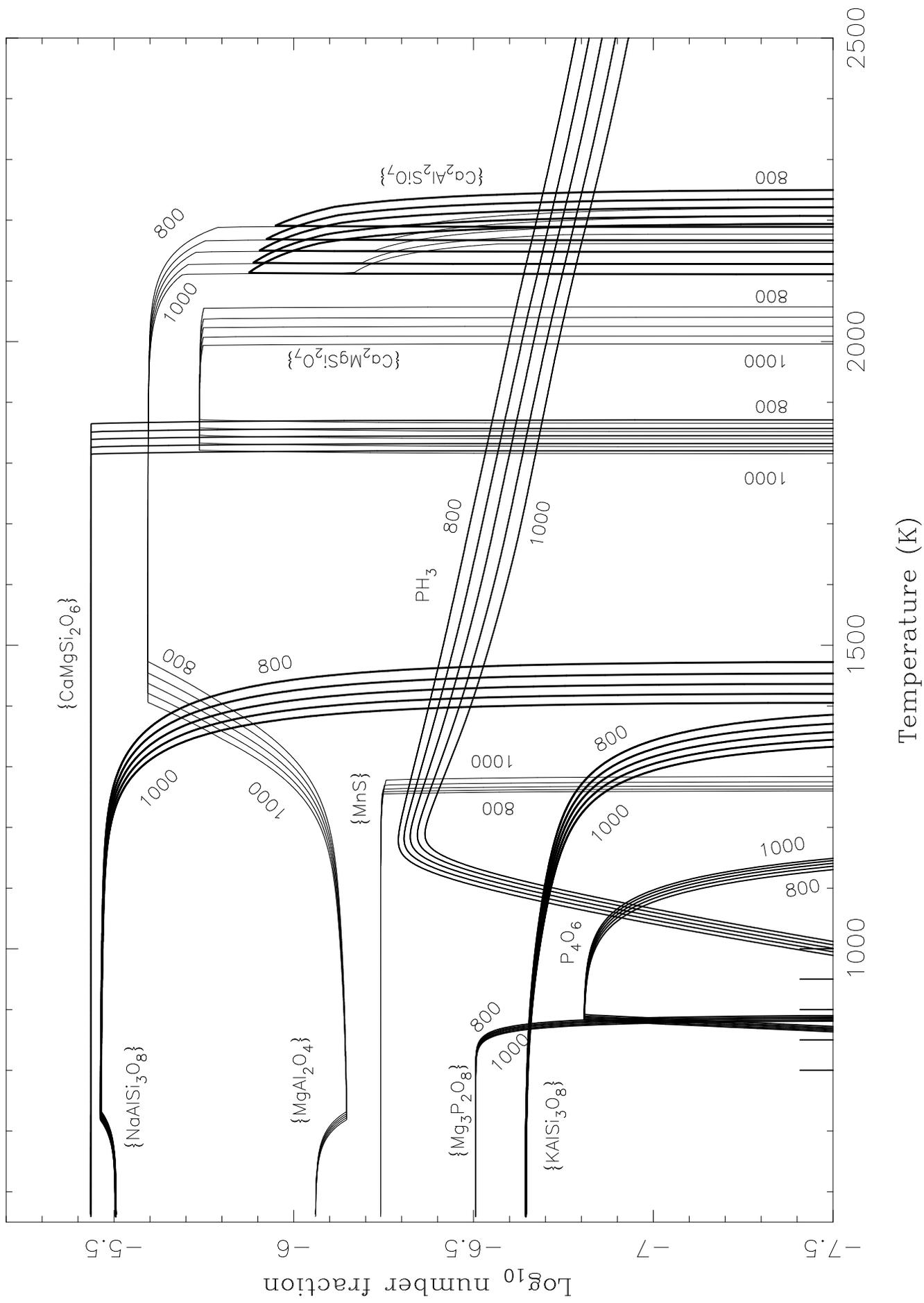
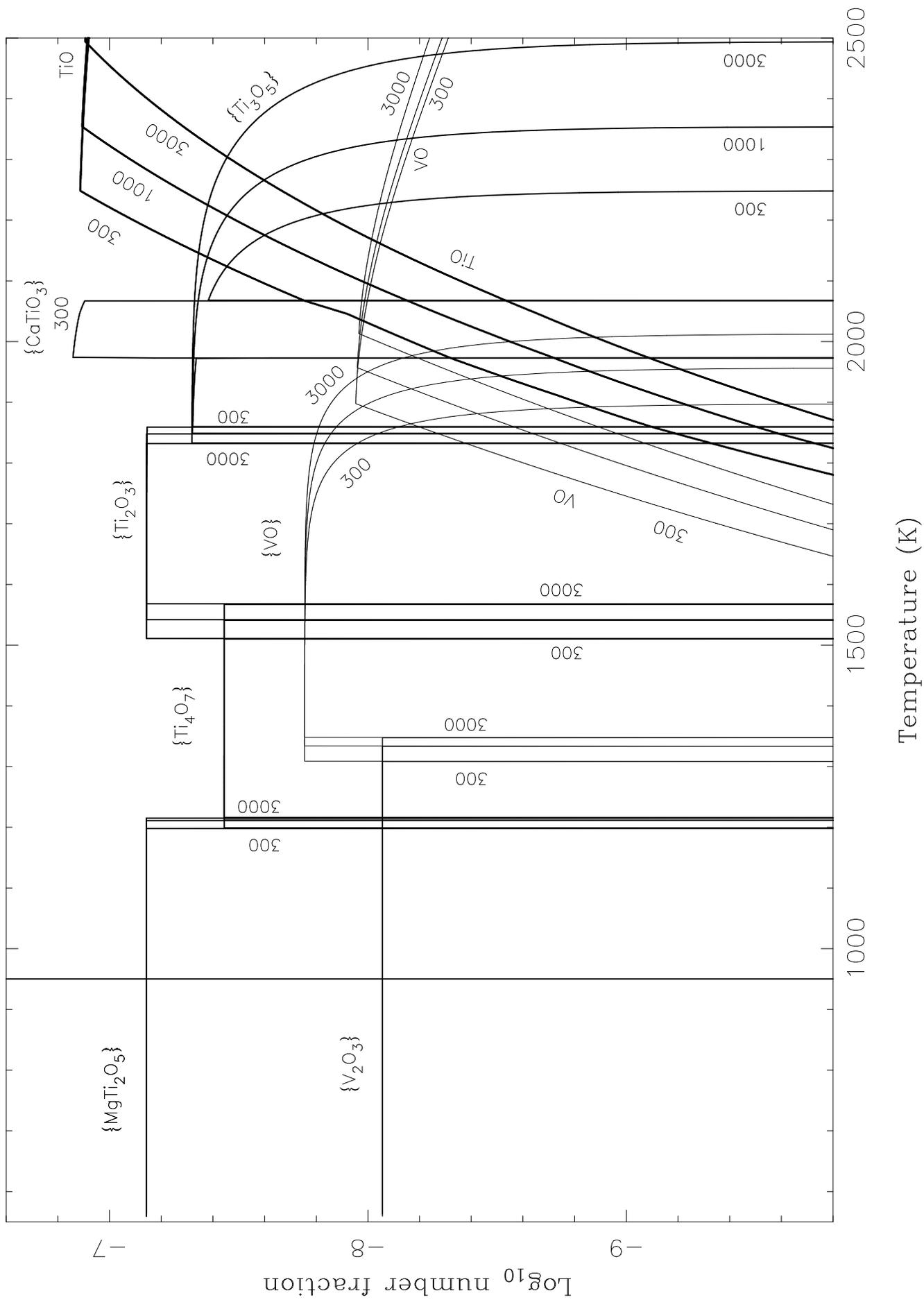
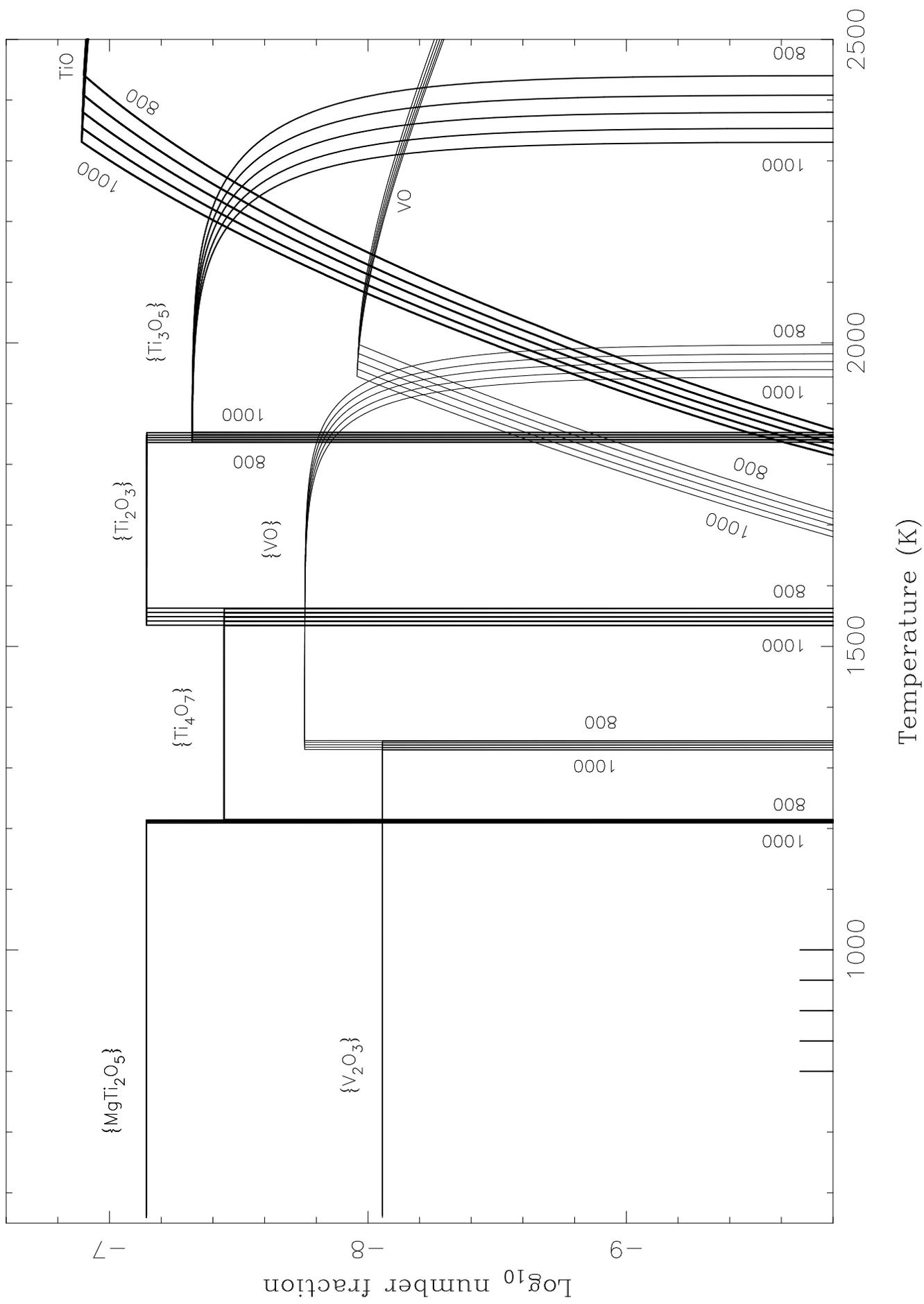
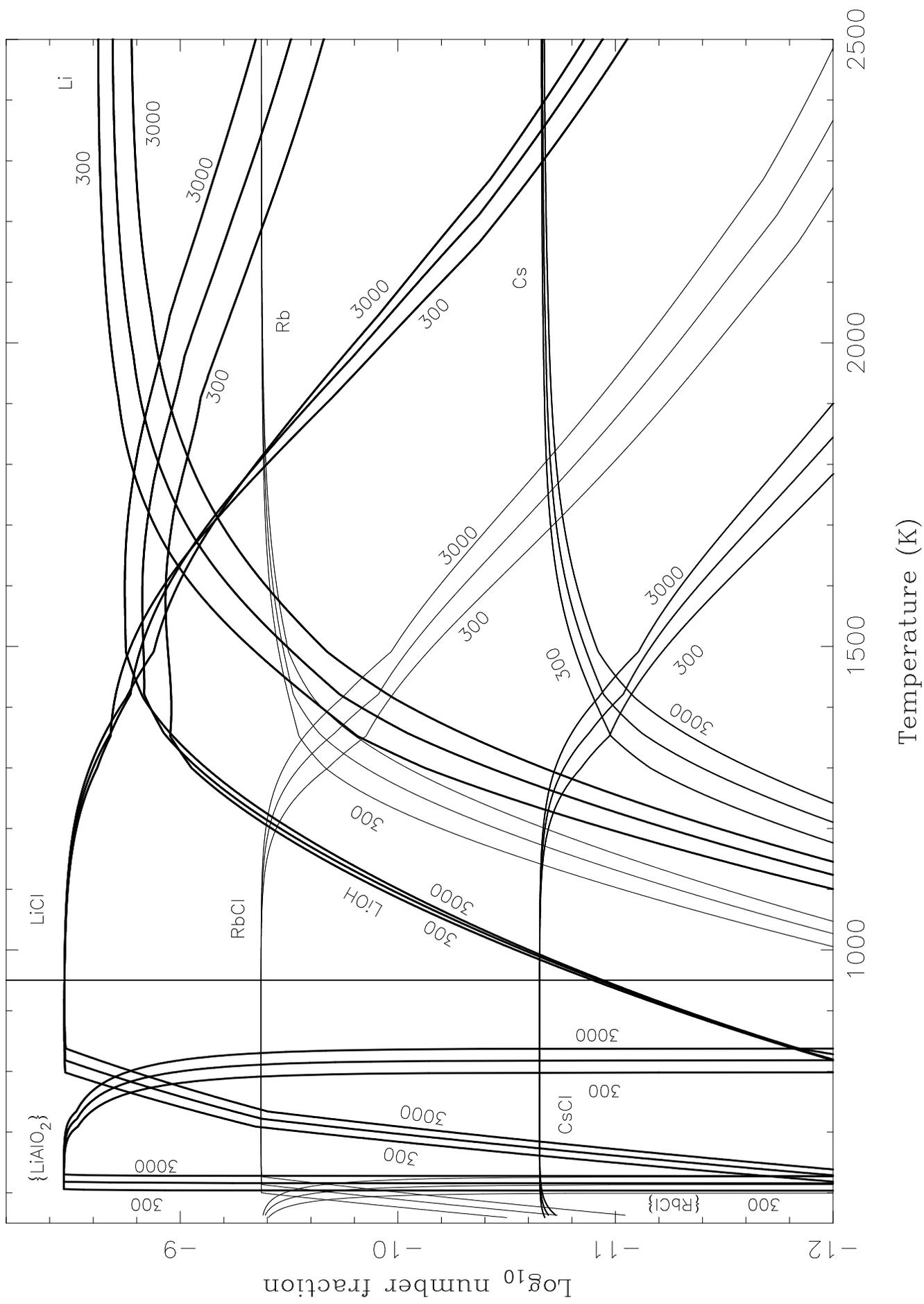
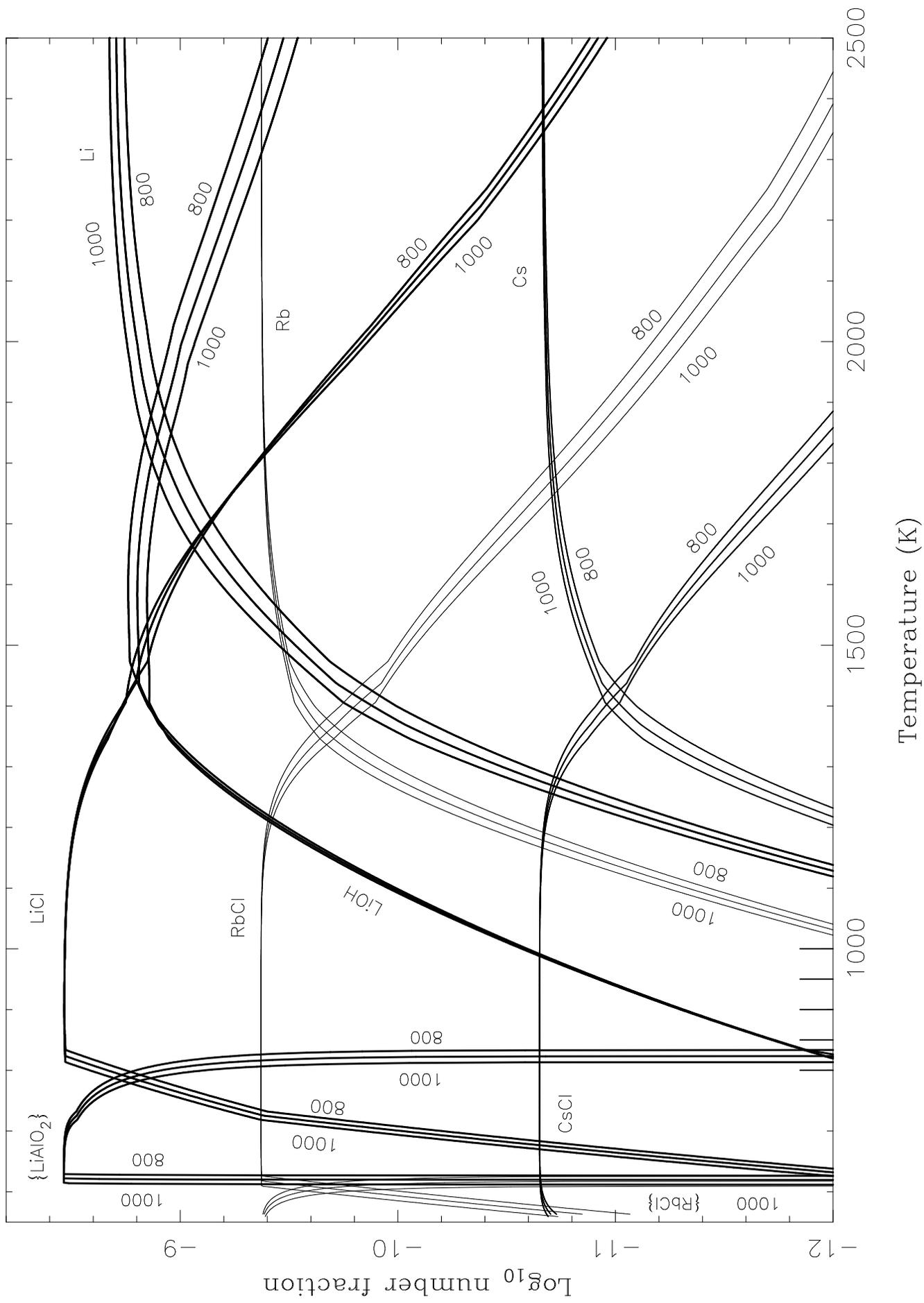
\end{document}